\documentclass[journal]{IEEEtran}
\usepackage{amsmath,amssymb}
\usepackage[dvips]{graphicx}
\usepackage{amsfonts}
\usepackage[mathscr]{eucal}
\usepackage{latexsym}
\usepackage{amsthm}
\usepackage{exscale}
\usepackage[mathscr]{eucal}
\usepackage{bm}
\usepackage[dvipsnames]{color}
\usepackage{cases}
\usepackage{epsfig}
\usepackage[center,small]{caption}
\usepackage{algorithm}
\usepackage{algorithmic}
\usepackage[verbose,nospace,sort]{cite}
\usepackage{tabularx}
\usepackage{multirow}
\usepackage{multicol}
\usepackage{balance}
\usepackage{bookmark}
\usepackage{amsmath}
\usepackage{amssymb}
\usepackage{mathrsfs}
\graphicspath{{./figs/}}

\usepackage{graphicx}

\usepackage[labelformat=brace]{subcaption}

\ifCLASSINFOpdf

\else

\fi

\scrollmode

\newtheorem{proposition}{Proposition}

\newtheorem{Lemma}{Lemma}

\begin{document}

\title{UAV-Enabled Radio Access Network: Multi-Mode Communication and Trajectory Design   }
\author{Jingwei~Zhang,
       Yong~Zeng,~\IEEEmembership{Member,~IEEE,}
        and~Rui~Zhang,~\IEEEmembership{Fellow,~IEEE}
\vspace{-0.7cm}
\thanks{J. Zhang and R. Zhang are with the Department of Electrical and
Computer Engineering, National University of Singapore (e-mail:~jingwei.zhang@u.nus.edu, elezhang@nus.edu.sg).}
\thanks{Y. Zeng is with the School of Electrical and Information Engineering, The University of Sydney, Australia 2006. He was with the Department of Electrical and Computer Engineering, National University of Singapore, Singapore 117583 (e-mail:yong.zeng@sydney.edu.au).}
}

\maketitle

\begin{abstract}
In this paper, we consider an unmanned aerial vehicle (UAV)-enabled radio access network (RAN) with the UAV acting as an aerial platform to communicate with a set of ground users (GUs) in a variety of modes of practical interest, including data collection in the uplink, data transmission in the downlink, and data relaying between GUs involving both the uplink and downlink. Under this general framework, two UAV operation scenarios are considered: \emph{periodic operation}, where the UAV serves the GUs in a periodic manner by following a certain trajectory repeatedly, and \emph{one-time operation} where the UAV serves the GUs with one single fly and then leaves for another mission. In each scenario, we aim to minimize the UAV periodic flight duration or mission completion time, while satisfying the target rate requirement of each GU via a joint UAV trajectory and communication resource allocation design approach. Iterative algorithms are proposed to find efficient locally optimal solutions by utilizing successive convex optimization and block coordinate descent techniques. Moreover, as the quality of the solutions obtained by the proposed algorithms critically depends on the initial UAV trajectory adopted, we propose new methods to design the initial trajectories for both operation scenarios by leveraging the existing results for solving the classic Traveling Salesman Problem (TSP) and Pickup-and-Delivery Problem (PDP). Numerical results show that the proposed trajectory initialization designs lead to significant performance gains compared to the benchmark initialization based on circular trajectory.
\end{abstract}


\section{Introduction}

To support the fast-growing traffic demand for the~next generation mobile communication systems, extensive research efforts have been devoted to exploring various new wireless technologies \cite{boccardi2014five}, such as ultra-dense network, millimeter wave (mmWave) communication, massive multiple-input multiple-output (M-MIMO), non-orthogonal multiple access (NOMA) \cite{7263349}, and machine-type communication. All these technologies were mainly developed for the terrestrial wireless network with base stations (BSs), relays and access points deployed at fixed locations. Recently, there have been significant interests in using unmanned aerial vehicles (UAVs) as aerial platforms to enable terrestrial communications from the sky \cite{zeng2016wireless}. Compared to conventional terrestrial communication, UAV-enabled communication is more swift and flexible to deploy for unexpected or temporary events. Besides, thanks to the UAV's high altitude, the favorable line-of-sight (LoS) communication links are more likely to be established between UAV and ground users (GUs) \cite{7108163}, \cite{khawaja2018survey}. Thus, UAV-enabled communication has many potential use cases, such as for public safety communication, ground BS offloading, emergency response, and Internet of things (IoT) communication.

Significant research efforts have been devoted to addressing the various new challenges for UAV-enabled communications, such as the UAV-ground channel characterization \cite{7108163,khawaja2018survey,7407385}, performance analysis \cite{5937283}, \cite{8254658}, and UAV placement optimization \cite{bor2016new,al2014optimal,lyu2016cyclical,lyu2017placement,8103781}. In particular, the controllable high mobility of UAVs offers a new design degree of freedom to enhance communication performance via trajectory optimization, which has received significant interests recently \cite{7572068,zeng2017energy,zeng2018energy,7932157,8247211,eom2018uav,8119562,8254971,7997208}.

In \cite{7572068}, the authors proposed a general framework via jointly optimizing the transmit power and UAV trajectory to maximize the end-to-end throughput for a UAV-enabled mobile relaying system. Specifically, the transmit power at the source/UAV relay and the UAV trajectory were optimized in an alternating manner iteratively via the technique of block coordinate descent. To tackle the non-convex trajectory optimization in each iteration, the successive convex optimization technique was proposed based on the local lower bound of the rate function. Such techniques have then been applied to various other scenarios in UAV-enabled wireless communications \cite{zeng2017energy,zeng2018energy,7932157,8247211,eom2018uav,8119562,8254971}. Note that for all these works employing successive convex optimization and block coordinate descent techniques, the converged results critically depend on the initial UAV trajectory adopted. A straight line based initial trajectory and a circular based initial trajectory were proposed in \cite{7572068} and \cite{8247211}, respectively. Though simple and intuitive, such trajectory initialization schemes do not fully exploit the locations and communication requirements of GUs. This thus gives one of the main motivations of the current work, to devise more sophisticated trajectory initialization schemes for UAV-enabled communications to achieve better converged performance.

It is worth noting that path planning or trajectory optimization has been extensively studied in the UAV control and navigation literature \cite{1023918,1184270,yu2015sense,tisdale2009autonomous,5518820,4288144}. For example, in \cite{1023918}, the UAV trajectory was formulated as a mixed integer~linear program (MILP) to ensure collision avoidance. In \cite{tisdale2009autonomous}, the receding-horizon path planning approach was applied to demonstrate the capability for a swarm of UAVs to perform autonomous search and localization. Moreover, the authors in \cite{5518820} and \cite{4288144} investigated the path planning for a single vehicle to collect data from all sensors. Note that the aforementioned works for path planning either focused on other design objectives rather than communication performance, or assumed simplified communication models, such as the disk model in \cite{5518820}, \cite{4288144}. In practice, adaptive communication with dynamic power and bandwidth allocation can be exploited along with the UAV trajectory design to achieve enhanced communication performance, as pursued in more recent works such as~\cite{7572068,zeng2017energy,zeng2018energy,7932157,8247211,eom2018uav}.

\begin{figure}
\vspace{-0.2cm}
\centering
\includegraphics[width=0.9\textwidth]{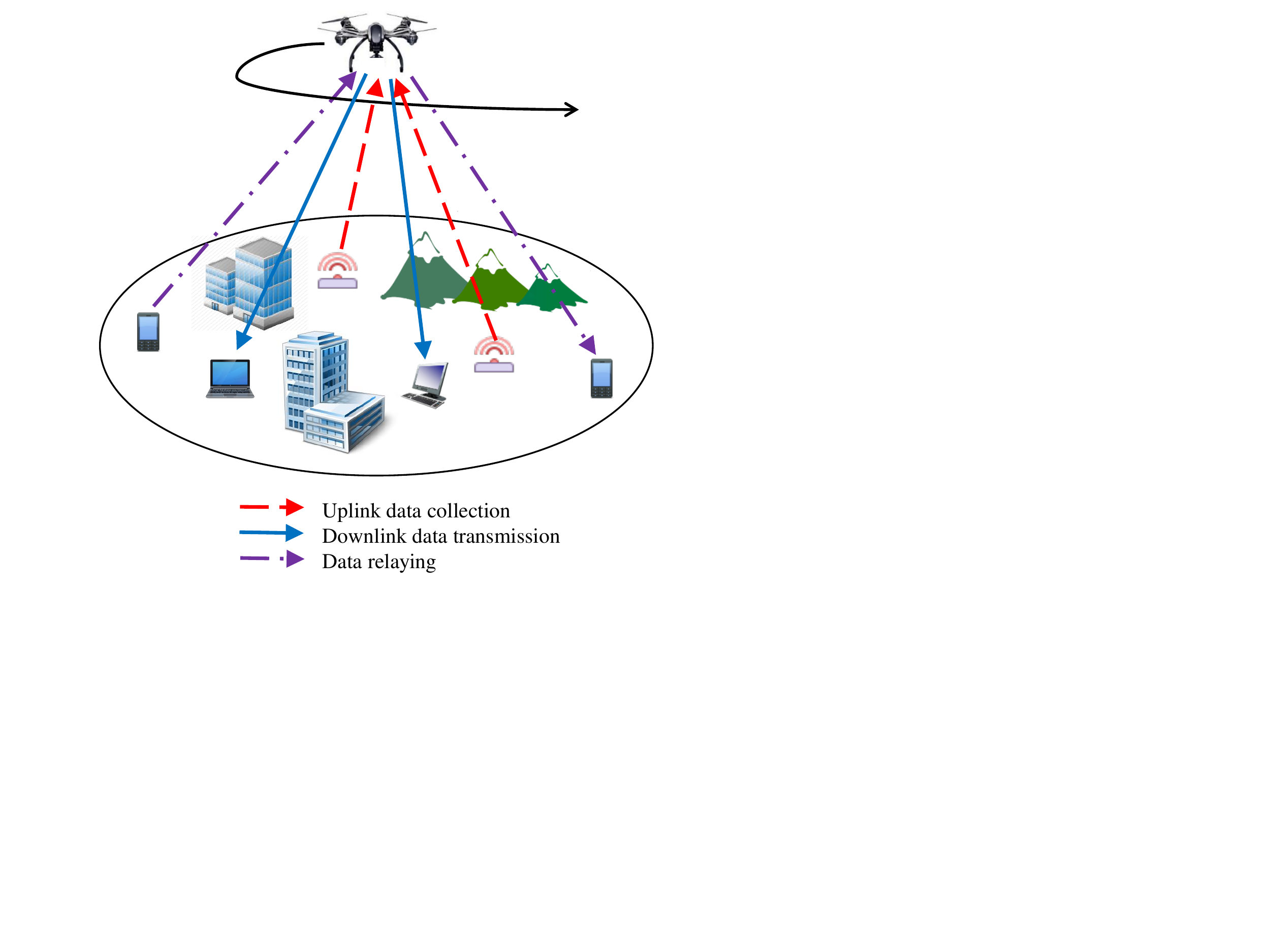}
\vspace{-5.5cm}
\caption{\label{system}A UAV-enabled aerial platform with multi-mode communications.}
\vspace{-0.65cm}
\end{figure}

In this paper, we study a general UAV-enabled radio access network (RAN) as shown in Fig. \ref{system}, where the UAV is employed as an aerial platform supporting multi-mode communications for its served GUs, including data relaying from one GU to another \cite{7572068}, downlink data transmission to GUs \cite{8247211}, and uplink data collection from GUs \cite{8119562} as special cases. Such a multi-mode aerial communication platform is more practically relevant for a real-life RAN with different traffic demands of the GUs.

For the considered general RAN, two UAV application scenarios of practical interest are further considered. The first one is {\it periodic operation}, where the UAV serves the GUs in a periodic manner by following a certain trajectory repeatedly. In this case, our objective is to minimize each periodic flight duration of the UAV for the purpose of minimizing the communication delay of the GUs \cite{lyu2016cyclical}, while satisfying the average rate requirement of each GU, via jointly optimizing the UAV trajectory, transmit power and bandwidth allocation. The second scenario corresponds to {\it one-time operation}, where the UAV serves the GUs with one single fly and then leaves for another mission. This may correspond to practical use cases such as periodic sensing, where the UAV only needs to be dispatched at a given frequency. In this scenario, we aim to minimize the mission completion time for saving UAV time for other missions while satisfying the aggregated throughput requirement of each GU, via jointly optimizing the UAV trajectory and pertinent communication resource allocation. In this case, for the particular data relaying mode, the UAV can only forward to a destination GU the data that has been received from its associated source GU, along its given one-round trajectory, thus resulting in a stringent {\it information-causality constraint} \cite{7572068}; whereas this constraint can be relaxed in the former periodic operation scenario thanks to the periodic trajectory of the UAV. The main contributions of this paper are summarized as follows.
\begin{itemize}
\item First, we propose a multi-mode UAV communication platform with periodic operation or one-time operation. For both operation scenarios, we formulate the optimization problems to minimize the UAV periodic flight duration and mission completion time, respectively, via jointly optimizing the UAV trajectory, bandwidth and power allocation. Since the formulated problems are difficult to be directly solved, we propose efficient iterative algorithms to find locally optimal solutions based on successive convex optimization and block coordinate descent techniques.

\item Second, as the converged results of the proposed algorithms critically depend on the initial UAV trajectory assumed, we propose new methods to design the initial trajectory by fully exploiting the location information and communication requirements of the GUs. Specifically, as the UAV typically has better communication link when it is near GUs, the initial UAV trajectory should be designed so as to approach each GU as much as possible. To this end, we propose the trajectory initialization design based on the Traveling Salesman Problem (TSP) solution for the case of periodic operation, and that based on the Pickup-and-Delivery Problem (PDP) solution for the case of one-time operation. Compared to the existing UAV initial trajectory designs such as the straight-line or circular trajectories, the main novelty of the proposed trajectory initialization lies in the optimized waypoints design and their order of visiting based on the number and location distribution of the GUs, their communication requirements as well as the UAV's practical mobility constraints such as its maximum speed.

\end{itemize}

The rest of this paper is organized as follows. Section \ref{sysmod} introduces the system model and presents the problem formulations for the periodic operation and the one-time operation scenarios, respectively.
Section \ref{period} and Section \ref{onetime} present the proposed algorithms based on successive convex optimization and block coordinate descent techniques for the two operation scenarios, respectively. In Section \ref{trajini}, we propose two efficient trajectory initialization designs for the two scenarios, respectively. Numerical results are presented in Section \ref{simulation} to evaluate the performance of the proposed designs. Finally, we conclude this paper in Section \ref{conclusion}.

{\it Notations:} In this paper, scalars and vectors are denoted by italic letters and boldface lower-case letters, respectively. $\mathbb{R}^{M\times 1}$ denotes the space of $M$-dimensional real-valued vectors. For a vector $\textbf{a}$, its Euclidean norm is represented by $||\textbf{a}||$. $\log_2(\cdot)$ denotes the logarithm with base 2. For a time-dependent function $\textbf{q}(t)$, $\dot{\textbf{q}}(t)$ represents the first-order derivative with respect to time $t$. For sets $\mathcal{M}_1$ and $\mathcal{M}_2$, $\mathcal{M}_1\cup \mathcal{M}_2$ means the union of the two sets.

\section{System Model and Problem Formulation}
\label{sysmod}
\subsection{System Model}

As shown in Fig. \ref{system}, we consider a general UAV-enabled wireless RAN, where a UAV serves as an aerial platform for a set of $K$ GUs. In general, the communication modes of the UAV-enabled wireless RAN can be classified into three categories as follows:

\subsubsection{Data Collection in Uplink}
The UAV is employed as a flying fusion center to collect data from GUs that are data sources on the ground, such as sensors in IoT \cite{8119562}.

\subsubsection{Data Transmission in Downlink}
In this mode, independent information is transmitted from the UAV to GUs. For example, the UAV may act as a data carrier with pre-cached data to transmit to the intended GUs \cite{xu2017overcoming}.

\subsubsection{Data Relaying}
The UAV functions as a mobile relay to assist in the communications between multiple pairs of GUs. For each pair, the data is firstly received from the source GU in the uplink and then forwarded to the destination GU in the downlink. By exploiting the LoS links between the UAV and GUs, UAV-enabled mobile relaying is a promising solution to overcome the unreliable terrestrial links between widespread GUs. Practical application scenarios include service recovery after natural disasters, emergency response, etc., \cite{5937283}, \cite{7122576}.

Accordingly, the GUs can be generally divided into three groups based on their communication modes. Group 1 corresponds to UAV-assisted data collection, which only involves the uplink communication from the GUs in this group to the UAV. Within this group, we assume that in total $K_1 \leq K$ independent information flows are transmitted from their respective GUs to the UAV. Group 2 corresponds to data transmission from the UAV to the GUs belonging to this group, where only downlink communication is involved and the UAV transmits in total $K_2\leq K$ independent information flows to their corresponding GUs. Lastly, Group 3 corresponds to data relaying, which involves both uplink and downlink communications. For this group, in total $K_3 \leq K$ information flows are firstly transmitted to the UAV from the source GUs in this group and then forwarded by the UAV to their respective destination GUs. For information relaying, we assume that the UAV employs the decode-and-forward (DF) strategy with a data buffer of sufficiently large size. Notice that in practice, we have $K\leq K_1+K_2+2K_3$, since each GU may correspond to multiple information flows. For ease of presentation, we assume that each GU is only associated with one information flow such that $K=K_1+K_2+2K_3$; whereas the developed results in this paper can be easily generalized to the cases with $K<K_1+K_2+2K_3$. By letting $U=K_1+K_3$ and $V=K_2+K_3$, we define a source GU set as $\mathcal{U}=\{1,\cdots,U\}$, with the first $K_3$ elements corresponding to source GUs from Group 3 (for information relaying) and the rest from Group 1 (for uplink data collection). Similarly, define a destination GU set as $\mathcal{V}=\{1,\cdots,V\}$ with the first $K_3$ GUs corresponding to destination GUs in Group 3 (for information relaying) and the rest from Group 2 (for downlink transmission). Without loss of generality, we assume that the source GU $k \in \mathcal{U} $ and the destination GU $k\in\mathcal{V}$, $k=1,\cdots,K_3$, correspond to the same pair in Group 3.

We consider a three-dimensional (3D) Cartesian coordinate system, where the locations of each source GU and destination GU are denoted as $\textbf{s}_i\in \mathbb{R}^{2\times 1}$, $i \in \mathcal{U}$, and $\textbf{d}_j\in \mathbb{R}^{2\times 1}$, $j\in \mathcal{V}$, respectively. We assume that the UAV flies at a given constant altitude $H$. Furthermore, for a given time horizon of duration $T$, denote the UAV trajectory projected on the ground as $\textbf{q}(t)\in \mathbb{R}^{2 \times 1}$, $0 \leq t \leq T$. Let $V_{\mathrm{max}}$ be the maximum UAV speed in meter/second (m/s). We then have the following constraint $||\dot{\textbf{q}}(t)|| \leq V_{\mathrm{max}}$, $0\leq t \leq T.$
The time-varying distance between the UAV and the GUs can be written as
\vspace{-0.2cm}
\begin{align}
\tilde{s}_{i}(t)=\sqrt{H^2+||\textbf{q}(t)-\textbf{s}_i||^2}, \ \ i\in \mathcal{U}, ~\!  \\
\tilde{d}_{j}(t)=\sqrt{H^2+||\textbf{d}_j-\textbf{q}(t)||^2}, \ \ j\in \mathcal{V}.
\vspace{-0.2cm}
\end{align}
We further assume that both the uplink and downlink channels are dominated by LoS links. Thereby, the channel power gains follow the free-space path loss model given by
\vspace{-0.1cm}
\begin{align}
h^u_{i}(t)=\lambda_0 \tilde{s}_{i}^{-2}(t), \ h^v_{j}(t)=\lambda_0 \tilde{d}_{j}^{-2}(t),\ \ \forall i,~j,
\end{align}
where $\lambda_0$ denotes the channel power gain at the reference distance of $\tilde{d}_0=1$ m.

Let the total available bandwidth be denoted as $B$. The UAV is assumed to employ the frequency division multiple access (FDMA) scheme with dynamic bandwidth allocation among all GUs. Specifically, at time instant $t$, denote $\alpha_i(t)$ as the fraction of the total bandwidth that is allocated for the source GU $i\in \mathcal{U}$, and $\beta_j(t)$ as that allocated for the destination GU $j\in \mathcal{V}$. We then have the following constraints:
\vspace{-0.2cm}
\begin{align}
\sum_{i=1}^U\alpha_i(t)+\sum_{j=1}^V\beta_j(t)\leq 1, \ \ \forall t,\\
\alpha_{i}(t)\geq 0,\ \
\beta_{j}(t)\geq 0, \ \ \forall i,j,t.\!\!
\end{align}
Note that the above dynamic FDMA scheme includes both conventional time division multiple access (TDMA) with dynamic user time scheduling and FDMA with fixed user bandwidth allocation as special cases. In particular, when all $\alpha_i(t)$ and $\beta_j(t)$ are set as binary values 1 or 0, we have the dynamic TDMA scheme. On the other hand, when $\alpha_i(t)=\alpha_i$, and $\beta_j(t)=\beta_j$, $\forall t$, we have the non-dynamic FDMA scheme.

Denote by $P_i^u$ the transmit power for the source GU $i \in \mathcal{U}$ if $\alpha_i(t)\neq 0$, which is assumed to be constant. The instantaneous normalized achievable rate in bits/second/Hertz (bps/Hz) for this GU can be expressed as
\vspace{-0.2cm}
\begin{align}
R_i^u(t)=\alpha_i(t) \log_2 \left(1+\frac{P_i^u h^u_i(t) }{B \alpha_i(t) N_0} \right)~~ ~~   \nonumber \\
=\alpha_{i}(t)\log_2 \left(1+\frac{P^u_{i}  \gamma_{i}(t)}{\alpha_{i}(t)}  \right), \ \ \forall i \in \mathcal{U}, \!\!\!\!\!\!\!\!\!\! \!\!
\end{align}
where $N_0$ represents the additive white Gaussian noise (AWGN) power spectral density in watts/Hz, and $\gamma_i(t)\triangleq \gamma_0/(H^2+||\textbf{q}(t)-\textbf{s}_i||^2)$ is the time-dependent channel-to-noise power ratio, and $\gamma_0\triangleq \lambda_0/(B N_0)$ denotes the reference signal-to-noise ratio (SNR) at the reference distance of $\tilde{d}_0=1$ m.

Similarly, let $p_j(t)$ denote the UAV's transmit power for the destination GU $j \in \mathcal{V}$ at time $t$. The instantaneous achievable rate in bps/Hz for this GU is thus expressed as
\vspace{-0.2cm}
\begin{align}
R_j^v(t)= \beta_j(t)\log_2\left(1+\frac{p_j(t) \gamma_0}{\beta_j(t)\tilde{d}_{j}^2(t) } \right) \nonumber~~~~ \\
=   \beta_j(t)\log_2\left(1+\frac{p_j(t) \rho_j(t)}{\beta_j(t)}    \right), \ \ \forall j \in \mathcal{V},   \!\!\!\!\!\!\!\!\! \!\!\!  \!\!\!\!\!
\end{align}
where $\rho_j(t)\triangleq \gamma_0/(H^2+||\textbf{d}_j-\textbf{q}(t)||^2 )$ is the channel-to-noise power ratio from the UAV to the destination GU $j$. Let $P^v$ denote the maximum total transmit power by the UAV. For the downlink transmission from the UAV to the $V$ destination GUs, we then have the following power constraint $\sum_{j=1}^V p_j(t) \leq P^v,~\forall t$.

\subsection{Problem Formulation}
Generally speaking, a UAV serving as a multi-mode aerial platform may have two operation scenarios in practice: {\it periodic operation} versus {\it one-time operation}, explained as follows.
\subsubsection{Periodic Operation}

With periodic operation, the UAV needs to remain airborne to serve the GUs periodically, where the GUs keep generating service requests to the UAV. We assume that the average rate requirements in bps for uplink and downlink communication corresponding to the different flows are $\bar{R}_i^u$, $i \in \mathcal{U}$, and $\bar{R}_j^v$, $j \in \mathcal{V}$, respectively. In particular, for the data relaying service in Group 3, the uplink and downlink average rate requirements for each pair should be balanced, i.e., $\bar{R}_k^u=\bar{R}_k^v$, $k=1,\cdots,K_3$. Without loss of generality, we assume that the UAV flies above the GUs following a periodic trajectory with period $T$, where $T$ is a design variable. Note that in practice, it is desirable to minimize $T$ in order to avoid large communication delay of GUs \cite{lyu2016cyclical}.

For notational convenience, define $\mathcal{Q} \triangleq \{\textbf{q}(t) \}$, $\mathcal{P}=\{p_j(t)\}$ and $\mathcal{B}=\{\alpha_i(t) , \beta_j(t) \}$. Our objective is to minimize the UAV flight period $T$, via jointly optimizing the UAV's trajectory $\mathcal{Q}$, the downlink transmit power $\mathcal{P}$, as well as the bandwidth allocation $\mathcal{B}$, while satisfying the average rate requirements by the GUs. The problem can be formulated as
\vspace{-0.2cm}
\begin{subequations}
\begin{align}
\mathrm{(P1)}      \min_{\substack{T, \mathcal{Q}, \mathcal{P}, \mathcal{B} }}     T    \nonumber  ~~~~~~~~~~~~~~~~~~~~~~~~\!~~~~\!~~~  \\
\label{p1001}
\mathrm{s.t.}~~\frac{B}{T}\int_0^T R_i^u(t) dt \geq  \bar{R}_i^u, \ \ \forall i \in \mathcal{U},~~  \\
\label{p1002}
\frac{B}{T} \int_0^TR_j^v(t)dt \geq \bar{R}_j^v, \ \ \forall j \in \mathcal{V},~~  \\
\label{p1003}
\sum_{j=1}^V p_j(t) \leq P^v, \ \ \forall t,~~~~~~~\!~~~~~\!~~~~\\
\label{p1004}
p_j(t)  \geq 0, \ \ \forall j,t,~~~~~~~~~~~~~~~~~~~\\
\label{p1005}
\sum_{i=1}^U\alpha_i(t)+\sum_{j=1}^V\beta_j(t)\leq 1, \ \ \forall t,\!\!\!\!\!\!\!\!\!\! ~~~~~\!~~~ \\
\label{p1006}
\alpha_{i}(t)\geq 0, \ \ \forall i,t,~~~~~~~\!~~~~~~~~~~~~~ \\
\label{p1007}
\beta_{j}(t)\geq 0, \ \ \forall j,t,~~~~~\!~\!~~~~~\!~~~~~~\!\!~~~~~\\
\label{p1008}
||\dot{\textbf{q}}(t)|| \leq V_{\mathrm{max}}, \ \ \forall t,~~~~~~~~~~~~~~~ \\
\label{p1009}
\textbf{q}(0)=\textbf{q}(T), ~~~~~~~~~~~~~~~~~~~~~~~~
\end{align}
\end{subequations}
where the constraint \eqref{p1009} ensures that the UAV returns to the initial location at the end of each period.

Different from the prior work \cite{8254949} which focuses on maximizing the minimum throughput over all GUs in downlink communication with given $T$, we here study the flight period minimization problem in a more general setup, where uplink communication, downlink communication and data relaying modes are all taken into account and $T$ is also a design variable.

\subsubsection{One-Time Operation}

In the second scenario, the UAV only needs to serve the GUs once by one single fly mission. This corresponds to many practical scenarios where the service requests by the GUs are intermittent. In this case, the UAV mission is regarded as completed once the throughput in bits (instead of the average rate as in periodic operation) for each information flow meets the target requirement of the GUs. Denote the uplink and downlink throughput requirements corresponding to different information flows as $C_i^u$ bits, $i \in \mathcal{U}$, and $C_j^v$ bits, $j\in \mathcal{V}$, respectively. Similar to the periodic operation scenario, for the particular data relaying service, the uplink and downlink throughput requirements should be balanced for each source-destination pair, namely  $C_k^u=C_k^v$, $k=1,\cdots,K_3$. Further, denote by $T$ the flight duration (or mission completion time) required by the UAV to meet the throughput requirements of all the information flows.

Furthermore, for data relaying in one-time operation scenario, we need to impose the stringent {\it information-causality constraints}, i.e., at any time instant $t$, the UAV can only forward the data that has already been previously received from the source GU $k$ in Group 3. Note that such information constraints do not need to be explicitly imposed for the periodic operation scenario since the UAV may forward the information received from the previous period, as long as the total information bits received from the source equal to that forwarded to the corresponding destination at each period to ensure the long-term balance. The {\it information-causality constraints} for data relaying in one-time operation can be expressed as \cite{7572068}
\vspace{-0.2cm}
\begin{align}
\label{infcau}
\int_0^t R_k^v(\tau) d\tau \leq \int_0^t R_k^u(\tau) d\tau,  \ \ k=1,\cdots,K_3, \forall t.
\end{align}
Note that the left-hand side (LHS) of \eqref{infcau} is the aggregated information bits that have been forwarded by the UAV to the destination GU $k$ at time $t$, and the right-hand side (RHS) represents those which have been received from the source GU $k$ at the same time. For one-time operation, we aim to minimize the mission completion time $T$ via a joint trajectory, spectrum and power allocation design. In practice, minimizing the completion time $T$ is of high practical interest since it helps save more time/energy for the UAV to serve other missions. The problem can be formulated as
\vspace{-0.2cm}
\begin{subequations}
\begin{align}
\mathrm{(P2)}      \min_{\substack{T, \mathcal{Q}, \mathcal{P}, \mathcal{B} }}     T    \nonumber  ~~~~~~~~~~~~~~~~~~~~~~~~~~~~  \\
\label{p2001}
\mathrm{s.t.}~~B\int_0^T R_i^u(t) dt \geq C^u_i, \ \ \forall i \in \mathcal{U}, \\
\label{p2002}
B\int_0^T R_j^v(t)dt \geq C^v_j , \ \ \forall j \in \mathcal{V}, \\
\eqref{p1003}-\eqref{p1008}, \ \ \eqref{infcau}. \nonumber~~~~~~~~~~~~~\!~~~
\end{align}
\end{subequations}
Note that in (P2), no constraints on the UAV's initial and final locations are imposed, i.e., they can be freely designed for performance optimization. The developed results can be easily extended to include such constraints similarly as in \cite{7572068}.

Besides, it should also be noted that in the prior work \cite{7572068}, the special case of UAV-enabled relaying with one pair of source and destination GUs has been studied, where the end-to-end throughput is maximized with a pre-determined time horizon $T$. In (P2), we study the completion time minimization problem in the general setup with multiple GUs and modes, where the results in \cite{7572068} cannot be directly applied.

\section{Proposed Solution for Periodic Operation}
\label{period}

In this section, we consider the flight period minimization problem (P1) for the periodic operation. Problem (P1) is challenging to solve for two reasons. First, the problem requires to optimize continuous functions $\mathcal{Q}$, $\mathcal{P}$ and $\mathcal{B}$, which essentially involve an infinite number of optimization variables that are closely coupled with each other. Secondly, the integral in the LHS of \eqref{p1001} and \eqref{p1002} involve the optimization variable $T$ as the upper bound of the integration interval, which lack closed-form expressions. To tackle these issues, we first introduce the following optimization problem for any given period $T$:
\vspace{-0.2cm}
 \begin{subequations}
\begin{align}
\mathrm{(P1.1)} \max_{\substack{\eta, \mathcal{Q}, \mathcal{P}, \mathcal{B} }}   \eta ~~~~~~~~~~~~~~~~~~~~~~ ~~~ \nonumber\\
\label{p11001}
\mathrm{s.t.}~~\frac{B}{T\bar{R}_i^u}\int_0^T R_i^u(t)dt \geq \eta, \ \ \forall i \in \mathcal{U}, \!\!\!\!\!\!\!\!\!\!\\
\label{p11002}
\frac{B}{T\bar{R}_j^v} \int_0^T R_j^v(t)dt  \geq \eta, \ \ \forall j \in \mathcal{V},\!\!\!\!\!\!\!\!\!\! \\
\eqref{p1003}-\eqref{p1009}.  \nonumber~~~~~~~~~~~~~~~~~~~~~
\end{align}
\end{subequations}
Problem (P1.1) aims to maximize the minimum ratio $\eta$ between the achievable average rate and the target rate requirement of each GU. For any given flight period $T$, denote the optimal value of (P1.1) as $\eta^*(T)$. It is not difficult to see that for any given $T$, the target rate requirements of all GUs are achievable if and only if $\eta^*(T)\geq 1$. Therefore, problem (P1) is equivalent to
\vspace{-0.2cm}
\begin{align} \mathrm{(P1.2)}  ~~    \min_{\substack{T~ }} ~~    T    \nonumber  ~~~~~~~~~~~~~~~~~~~~~~~~\\
\vspace{-0.1cm}
\label{p12001}
\vspace{-0.2cm}
\mathrm{s.t.}~~~\eta^*(T) \geq 1.~~~~~~~~~~~~~~~~~~~~
\end{align}
\begin{Lemma}
\label{lemma001}
The optimal value $\eta^*(T)$ of problem (P1.1) is non-decreasing with $T$.
\end{Lemma}
\vspace{-0.2cm}
\begin{IEEEproof}
Please refer to Appendix \ref{proflema01}.
\end{IEEEproof}

By applying Lemma \ref{lemma001}, problem (P1.2) can be solved by applying a bisection search over $T$ until the equality in \eqref{p12001} holds. Thus, the main task of solving (P1) is to find an efficient algorithm for (P1.1) for any given $T$.


To obtain a more tractable form of (P1.1), we apply a discrete state-space approximation. Specifically, the time horizon $T$ is equally divided into $N$ time slots, i.e., $t_n= n \delta_t$, $n=1,\cdots,N$, with $\delta_t$ representing the time step which is sufficiently small such that the distance between the UAV and the GUs can be assumed to be approximately constant within each time slot. Therefore, the UAV's trajectory $\textbf{q}(t)$ over $T$ can be specified by $\textbf{q}[n]\triangleq \textbf{q}(n \delta_t)$, $n=1,\cdots,N$. As a result, the UAV speed constraints \eqref{p1008} can be represented as
\begin{align}
||\textbf{q}[n+1]-\textbf{q}[n]||^2\leq D_{\mathrm{max}}^2   , \ \ n=1,\cdots, N-1,
\end{align}
where $D_{\mathrm{max}} \triangleq V_{\mathrm{max}}\delta_t$ denotes the maximum distance that the UAV can travel within each time slot. The bandwidth and transmit power allocation can be similarly discretized as $\alpha_i[n] \triangleq\alpha_i(n\delta_t)$, $\beta_j[n]\triangleq\beta_j(n\delta_t)$, $p_j[n]\triangleq p_j(n \delta_t)$, $\forall i,j, n$. Then, the achievable rate between the GUs and the UAV at time slot $n$ can be expressed as
\vspace{-0.22cm}
\begin{align}
R_i^u[n]=\alpha_{i}[n]\log_2 \left(1+\frac{P^u_{i}  \gamma_{i}[n]}{\alpha_{i}[n]}  \right), \ \ \forall i,n,~~~\!\!\!\\
R_j^v[n]=\beta_{j}[n]\log_2 \left(1+\frac{p_{j}[n] \rho_{j}[n] }{\beta_{j}[n]}   \right), \ \  \forall j, n,\!
\end{align}
where
\vspace{-0.2cm}
\begin{align}
\label{gamma}
\gamma_{i}[n]\triangleq \frac{\gamma_0}{ H^2+||\textbf{q}[n]- \textbf{s}_i||^2 }, \\
\label{rho}
\rho_{j}[n]\triangleq\frac{\gamma_0}{H^2+||\textbf{d}_j-\textbf{q}[n]||^2}.  \!\!
\end{align}

Besides, $\mathcal{Q}$, $\mathcal{P}$ and $\mathcal{B}$ are rewritten as $\mathcal{Q} \triangleq \{\textbf{q}[n], \forall n \}$, $\mathcal{P}=\{p_j[n], \forall j,n \}$ and $\mathcal{B}=\{\alpha_i[n] , \beta_j[n], \forall i,j,n  \}$, respectively. As a result, problem (P1.1) is reformulated as
\vspace{-0.2cm}
\begin{subequations}
\begin{align}
\!\!\!\mathrm{(P1.3)} \max_{\substack{\eta, \mathcal{Q}, \mathcal{P}, \mathcal{B} }}   \eta ~~~~~~~~~~~~~~~~~~~~~~~~~~~~~~~~~~~~~~ ~~~~~~ \nonumber\\
\label{p13001}
\mathrm{s.t.}~\frac{B}{N \bar{R}_i^u}\sum_{n=1}^{N} R_i^u[n] \geq \eta, \ \ \forall i \in \mathcal{U}, ~~~~~~~~~~~~~~~~~~~~\\
\label{p13002}
\frac{B}{N \bar{R}_j^v}\sum_{n=1}^{N} R_j^v[n] \geq \eta, \ \ \forall j \in \mathcal{V},~~~~~~~~~~~~\!~~~~~~~~\\
\label{p13003}
\sum_{j=1}^V p_j[n]\leq P^v, \ \ \forall n, ~~~~~\!~~~~~~~\!~~~~~~~~~~~~~~~~~~~~\!\\
\label{p13004}
p_j[n]\geq 0, \ \ \forall j, n,~~~~~~~~~~~~~~~~~~~~~~~~ ~~~\!~~~~~~~~\\
\label{p13005}
\sum_{i=1}^U \alpha_i[n]+\sum_{j=1}^V \beta_j[n] \leq 1, \ \  \forall n,~~~~~~~~~~~~~\!~~~~~~\\
\label{p13006}
\alpha_i[n] \geq 0, \ \ \forall i,n,~~~~~~~~~~~~~~~~~~~~~~~~~~~~~~~~~~~\\
\label{p13007}
\beta_j[n] \geq 0, \ \ \forall j,n,~~~~~~~~~~~~~~~~~~~~~~~~~~~~~~~~~~~\\
\label{p13008}
||\textbf{q}[n+1]-\textbf{q}[n]||^2\leq D_{\mathrm{max}}^2, \ \ n=1,\cdots, N-1,\!\!\!\!\\
\label{p13009}
\textbf{q}[1]=\textbf{q}[N],~~~~~~~~~~~~~~~~~~~~~~~~~~~~~~~~~~~~~~~~~
\end{align}
\end{subequations}
where constraints \eqref{p13001}-\eqref{p13009} represent the discrete-time equivalents of \eqref{p11001}, \eqref{p11002}, \eqref{p1003}-\eqref{p1009}, respectively. As constraints \eqref{p13001} and \eqref{p13002} are non-convex with respect to variables $\mathcal{Q}$, $\mathcal{P}$ and $\mathcal{B}$, problem (P1.3) is difficult to be directly solved in general. In the following, we propose an efficient suboptimal solution to (P1.3) based on successive convex optimization and block coordinate descent techniques, similarly as in \cite{7572068}. The main idea is to solve the two sub-problems of (P1.3) iteratively, namely the power and bandwidth optimization with fixed trajectory, and trajectory optimization with fixed power and bandwidth allocation. Then, the block coordinate descent method is employed to optimize the two sets of variables in an alternating manner until the objective value $\eta$ converges within a prescribed accuracy.

\subsection{Power and Bandwidth Optimization with Fixed Trajectory}

First, we consider the sub-problem to optimize the UAV transmit power $\mathcal{P}$ and bandwidth allocation $\mathcal{B}$, for any given feasible UAV trajectory $\mathcal{Q}$. In this case, the time-varying variables $\{\gamma_{i}[n],\rho_{j}[n]\}_{n=1}^N$ in \eqref{gamma} and \eqref{rho} are also determined. This sub-problem of (P1.3) is given by
\vspace{-0.2cm}
\begin{subequations}
\begin{align}
\mathrm{(P1.4)} ~~ \max_{\substack{\eta, \mathcal{P}, \mathcal{B} }}  ~~ \eta ~~~~~~~~~~~~~~~~~~~~~~~~~~~~~~~~~~~~~~~~~~\nonumber \\
\label{p14001}
\!\!\!\!\!\!\!\!\!\mathrm{s.t.}~\frac{B}{N \bar{R}_i^u} \sum_{n=1}^{N} \alpha_i[n]\log_2 \left(1+\frac{P_i^u \gamma_i[n]}{\alpha_i[n]} \right) \geq \eta,\ \  \forall i \in \mathcal{U}, \!\!\!\!   \\
\label{p14002}
\!\!\!\frac{B}{N \bar{R}_j^v}  \sum_{n=1}^N \beta_j[n] \log_2 \left(1+\frac{p_j[n]\rho_j[n]}{\beta_j[n]} \right) \geq \eta, \ \ \forall j \in \mathcal{V},\!\! \!  \\
\eqref{p13003}-\eqref{p13007}.  \nonumber~~~~~~~~~~~~~~~~~~~~~~~~~~~~~~~~~~~~~~~~~~
\end{align}
\end{subequations}
It can be shown that the LHS of \eqref{p14001} is concave with respect to the bandwidth allocation $\alpha_i[n]$, and the LHS of \eqref{p14002} is jointly concave with respect to the bandwidth allocation $\beta_j[n]$ and the transmit power $p_j[n]$, and all other constraints are convex. Therefore, (P1.4) is a convex optimization problem, which can be efficiently solved via existing software such as CVX \cite{grant2008cvx} or applying the Lagrange duality \cite{boyd2004convex}, for which the details are omitted for brevity.

\subsection{Trajectory optimization with Fixed Power and Bandwidth Allocation}

In this subsection, we consider the other sub-problem to optimize the UAV trajectory $\mathcal{Q}$ by assuming that the transmit power $\mathcal{P}$ and bandwidth allocation $\mathcal{B}$ are given. However, even with fixed power and bandwidth allocation, the trajectory optimization in (P1.3) is still a non-convex problem due to non-convex constraints \eqref{p13001} and \eqref{p13002}. To tackle such non-convexity, the successive convex optimization technique similar to that used in \cite{7572068} and \cite{zeng2017energy} can be applied, for which a lower bound of the original problem is sequentially maximized by optimizing the trajectory at each iteration. To this end, we need the following result.\vspace{-0.2cm}
\begin{proposition}
\label{theorem01}
For any given local trajectory $\mathcal{Q}^l \triangleq \{\textbf{\emph{q}}^l[n], \forall n\}$, we have
\vspace{-0.4cm}
\begin{align}
\label{ruk}
\!\!\!\!\!\!R^u_{i}[n] \geq \hat{R}^u_{i}[n]
\triangleq   \alpha_{i}[n] \log_2 \left(1+\frac{\varepsilon_{i}[n]}{H^2+||\textbf{\emph{q}}^{l}[n] -\textbf{\emph{s}}_i||^2} \right)\!\!\! \!\!\!\!\!\!\!\! \nonumber\\
-\phi_{i}^{l}[n]\left( ||\textbf{\emph{q}}[n]-\textbf{\emph{s}}_i ||^2-||\textbf{\emph{q}}^l[n]-\textbf{\emph{s}}_i ||^2 \right), \ \ \forall i,n,\!\! \!\!\\
\label{rdk}
\!\!\!\!\!\!R^v_{j}[n] \geq \hat{R}_{j}^{v} [n]
\triangleq  \beta_{j}[n] \log_2\left(1+\frac{\zeta_{j}[n]}{ H^2+||\textbf{\emph{d}}_j-\textbf{\emph{q}}^{l}[n]||^2} \right)\!\!\! \!\!\!\!\!\!\!\!\!\nonumber\\
-\varphi_{j}^{l}[n]\left( ||\textbf{\emph{d}}_j -\textbf{\emph{q}}[n] ||^2-||\textbf{\emph{d}}_j-\textbf{\emph{q}}^l[n] ||^2 \right),    \ \ \forall  j, n,\!\!\!\!\!\!\!\!~\!\!\!
\end{align}
where $\varepsilon_{i}[n]$$\triangleq$$ P^u_{i}\gamma_0/\alpha_{i}[n]$, $\zeta_{j}[n]$$\triangleq$$p_{j}[n]\gamma_0/\beta_{j}[n]$, coefficients $\phi_{i}^{l}[n]$ and $\varphi_{j}^l[n]$ are given in Appendix \ref{proftheorem01}. Both inequalities in \eqref{ruk} and \eqref{rdk} are active at $\textbf{\emph{q}}[n]=\textbf{\emph{q}}^l[n]$, $\forall n$.
\end{proposition}
\begin{IEEEproof}
Please refer to Appendix \ref{proftheorem01}.
\end{IEEEproof}

Proposition \ref{theorem01} shows that for any given local trajectory $\textbf{q}^l[n]$, $R^u_{i}[n]$ and $R^v_{j}[n]$ are respectively lower-bounded by $\hat{R}_{i}^{u} [n]$ and $\hat{R}_{j}^{v} [n]$, which are both concave functions with respect to $\textbf{q}[n]$. As a result, for any given local trajectory  $\mathcal{Q}^l$, a lower bound of the optimal value of the original problem (P1.3) with fixed power and bandwidth allocation can be obtained by solving the following problem
\vspace{-0.2cm}
\begin{subequations}
\begin{align}
\mathrm{(P1.5)} ~~\max_{\substack{\eta, \mathcal{Q}  }}~~   \eta~~~~~~~~~~~~~~~~~~~~~~~~~~~~\nonumber\\
\label{p16001}
~~~\mathrm{s.t.}~~~\frac{B}{N \bar{R}_i^u}\sum_{n=1}^{N} \hat{R}^{u}_{i}[n] \geq \eta, \ \ \forall i \in \mathcal{U},~
\end{align}
\begin{align}
\label{p16002}
\frac{B}{N \bar{R}_j^v}\sum_{n=1}^{N} \hat{R}^{v}_{j}[n] \geq \eta, \ \ \forall j \in \mathcal{V}, ~\\
\eqref{p13008},~\eqref{p13009}. ~~~~ ~~~~~~~~~~~~\nonumber~~~~~~~
\end{align}
\end{subequations}
Note that due to the lower bound given in Proposition \ref{theorem01}, if \eqref{p16001} and \eqref{p16002} are satisfied, then the constraints \eqref{p13001} and \eqref{p13002} with the same power and bandwidth allocation are guaranteed to be satisfied as well, but the reverse is not true. Therefore, the feasible region of (P1.5) is in general a subset of that of (P1.3), and its optimal solution serves as a lower bound to that for (P1.3) with fixed power and bandwidth allocation. (P1.5) is a convex optimization problem, which can be efficiently solved with the standard convex optimization techniques or existing solvers such as CVX \cite{grant2008cvx}.

\subsection{Iterative Power, Bandwidth and Trajectory Optimization}

Based on the results obtained above, we propose an iterative algorithm for (P1.3) based on the block coordinate descent technique. The details are summarized in Algorithm \ref{iter1}.
\vspace{-0.2cm}
\begin{algorithm}
\caption{Iterative power, bandwidth and trajectory optimization for (P1.3).\label{iter1}}
1: ~Initialize the UAV's trajectory as $\mathcal{Q}^l$ and let $l=0$.\\
2: ~~~\textbf{repeat}\\
3:~~~~~~For given $\mathcal{Q}^l$, obtain the optimal power and bandwidth \\
 \hspace*{0.8cm} allocation $\mathcal{P}^{l+1}$, $\mathcal{B}^{l+1}$ by solving (P1.4). \\
4:~~~~~~\!~~\!\!\!For given $\mathcal{P}^{l+1}$, $\mathcal{B}^{l+1}$ as well as $\mathcal{Q}^l$, update the UAV's \\
 \hspace*{0.85cm}  trajectory $\mathcal{Q}^{l+1}$ by solving (P1.5). \\
5:~~~\!~~~Update $l=l+1$.\\
6: ~~~\textbf{until} $\eta$ converges within a prescribed accuracy or a\\
 \hspace*{.65cm} maximum  number of iterations has been reached.
\end{algorithm}
\vspace{-0.2cm}

Since in each iteration of Algorithm \ref{iter1}, (P1.5) is optimally solved with given local trajectory $\mathcal{Q}^l$, whose objective value is non-decreasing over iterations and upper-bounded by a finite value, Algorithm \ref{iter1} is guaranteed to converge to at least a locally optimal solution. Note that for step 4 of Algorithm \ref{iter1}, an alternative way is to successively optimize the trajectory multiple times until convergence. The resulted objective value is also non-decreasing over iterations, thus its convergence is also guaranteed.



\section{Proposed Solution for One-Time Operation}
\label{onetime}

In this section, we study the optimization problem (P2) for the one-time operation. Similar to (P1), in order to solve (P2), we first consider the following problem for any given UAV operation time $T$:
\vspace{-0.2cm}
\begin{subequations} \begin{align}
\mathrm{(P2.1)}~\max_{\substack{\eta, \mathcal{Q}, \mathcal{P}, \mathcal{B} }} ~~  \eta~~~~~~~~~~~~~~~~~~~~~~~~~\nonumber \\
\label{p21001}
\mathrm{s.t.}~~\frac{B}{C^u_i}\int_{0}^{T} R_i^u(t) dt \geq \eta, \ \ \forall i \in \mathcal{U},\\
\label{p21002}
~~\frac{B}{C^v_j}\int_{0}^{T} R_j^v(t) dt \geq \eta, \ \ \forall j \in \mathcal{V}, \\
\eqref{p1003}-\eqref{p1008}, \ \ \eqref{infcau}. \nonumber~~~~~~~\!~~~~~~\!~\! ~~~~
\end{align}
\end{subequations}

Problem (P2.1) aims to maximize the minimum ratio $\eta$ between the achievable throughout and the target requirement. For any given operation time $T$, let the optimal solution to (P2.1) be denoted as $\eta^*(T)$. Then it is not difficult to see that all throughput requirements of (P2) are achievable if and only if $\eta^*(T)\geq 1$. Therefore, (P2) is equivalent to finding the minimum $T$ such that $\eta^*(T)\geq 1$. Furthermore, as the time $T$ only appears in the upper limit of the integral in \eqref{p21001} and \eqref{p21002} (no normalization by $T$ as in (P1.1)), it is quite obvious that the LHS of \eqref{p21001} and \eqref{p21002}  are non-decreasing with $T$. Thus, $\eta^*(T)$ is also non-decreasing with $T$. Therefore, (P2) can be solved by solving (P2.1) and applying a bisection search over the completion time $T$.

Similar to Section \ref{period}, for any given $T$, problem (P2.1) can be recast in a discrete equivalent form as
\vspace{-0.2cm}
\begin{subequations} \begin{align}
\!\!\!\!\!\!\!\!\!\!\!\!\!\!\mathrm{(P2.2)}~\max_{\substack{\eta, \mathcal{Q}, \mathcal{P}, \mathcal{B} }} ~~  \eta~~~~~~~~~~~~~~~~~~~~~~~~~~~~~~\nonumber \\
\label{p22001}
\mathrm{s.t.}~~\frac{B\delta_t}{C^u_i}\sum_{n=1}^{N-1} R_i^u[n] \geq \eta, \ \ \forall i\in \mathcal{U},~~~~~~\\
\label{p22002}
\frac{B\delta_t}{C^v_j}  \sum_{n=2}^N R_j^v[n] \geq \eta, \ \ \forall j \in \mathcal{V},~~~~~~\\
\label{p22003}
\sum_{m=2}^{n}R^v_{k}[m] \leq \sum_{m=1}^{n-1} R^u_{k}[m],~~~~~~~~~~ \nonumber  \\
k=1,\cdots,K_3, n=2,\cdots,N,  \!\!\!\!\!\!\!\!\!\!\!\!\!~~\!\!\!\!\!\!\!\!\!\!\!  \\
\eqref{p13003}-\eqref{p13008},~~~~~~~~~~~~~~~~~~~~~~~~  \nonumber
\end{align}
\end{subequations}
where \eqref{p22003} represents the discrete-time equivalent of the {\it information-causality constraints} in \eqref{infcau}. As constraints \eqref{p22001}-\eqref{p22003} are non-convex, problem (P2.2) is difficult to be optimally solved. Similar to Section \ref{period}, we apply the successive convex optimization and block coordinate descent techniques to (P2.2) by iteratively solving the two sub-problems, namely the power and bandwidth optimization with fixed trajectory, and trajectory optimization with fixed power and bandwidth allocation, as detailed in the next.

\vspace{-0.4cm}
\subsection{Power and Bandwidth Optimization with Fixed Trajectory}
With the given UAV trajectory $\mathcal{Q}$, problem (P2.2) reduces~to optimizing the UAV transmit power $\mathcal{P}$ and bandwidth allocation $\mathcal{B}$. By introducing slack variables $\{R_k^r[n] \}_{n=2}^N$, $k=1,\cdots,K_3$, problem (P2.2) can be equivalently transformed to
\vspace{-0.2cm}
\begin{subequations}
\begin{align}
\mathrm{(P2.3)} \max_{\substack{\eta, \{R^r_{k}[n] \},  \mathcal{P}, \mathcal{B}  }}   \eta \nonumber~~~~~~~~~~~~~~~~~~~~~~~~~~\!~~~~~  ~~~~~\\
\label{p23001}
\!\!\!\!\!\!\mathrm{s.t.}~~ \frac{B\delta_t}{C^u_i} \sum_{n=1}^{N-1}\alpha_{i}[n]\log_2\left(1+\frac{P^u_{i}\gamma_i[n]}{\alpha_{i}[n]} \right)\geq \eta, \ \forall   i \in \mathcal{U},\!\!\! \!\!\!\\
\frac{B\delta_t}{C^v_j}\sum_{n=2}^N\beta_j[n]\log_2\left(1+\frac{p_j[n]\rho_j[n]}{\beta_{j}[n]} \right)\geq \eta, \ \ ~~~\nonumber\\
j=K_3+1,\cdots,V,   \\
\label{p23002}
\frac{B\delta_t}{C^v_k} \sum_{n=2}^{N}R^r_{k}[n] \geq \eta, \ \  k=1,\cdots,K_3, ~~~~~~~~~~~\\
\label{p23003}
R^r_{k}[n]\leq \beta_{k}[n] \log_2\left(1+\frac{p_k[n]\rho_k[n]}{\beta_{k}[n]} \right), ~\nonumber  ~~~~~~~\!~\\
k=1,\cdots,K_3, n=2,\cdots,N,
\end{align}
\begin{align}
\label{p23004}
\sum_{m=2}^n R^r_{k}[m] \leq \sum_{m=1}^{n-1} \alpha_{k}[m]\log_2\left(1+\frac{P^u_{k}\gamma_k[m]}{\alpha_{k}[m]} \right), \!\!\!\! \nonumber
\\ k=1,\cdots,K_3,  n=2,\cdots,N, \\
\eqref{p13003}-\eqref{p13007}. \nonumber~~~~~~~~~~~~~~~~~~~~~~~~~~~~~~~~~~~~~~
\end{align}
\end{subequations}
Note that if at the optimal solution to (P2.3), there exists one constraint in \eqref{p23003} that is satisfied with strict inequality, we are always able to decrease the corresponding transmit power $p_k[n]$ and/or the bandwidth allocation $\beta_k[n]$ to make~the constraint active. This implies that there always exists an optimal solution to (P2.3) at which all constraints in \eqref{p23003} are active, and thus (P2.3) is equivalent to (P2.2) for any given trajectory. Furthermore, it can be verified that all constraints of (P2.3) are convex, thus (P2.3) is a convex optimization problem, which can be efficiently solved via standard convex optimization software such as CVX \cite{grant2008cvx}.

\subsection{Trajectory optimization with Fixed Power and Bandwidth Allocation}

Next, we consider the other sub-problem to optimize the UAV trajectory $\mathcal{Q}$ for any given transmit power $\mathcal{P}$ and bandwidth allocation $\mathcal{B}$. To deal with non-convex constraints \eqref{p22001}-\eqref{p22003}, the successive convex optimization is employed based on the lower bounds given in Proposition \ref{theorem01}. Specifically, for any given local trajectory, by introducing slack variables $\{R_k^r[n] \}$, the resulted problem is given by
\vspace{-0.2cm}
\begin{subequations}
\begin{align}
\mathrm{(P2.4)} \max_{\substack{\eta, \{R^r_{k}[n] \}, \mathcal{Q}  }}~~   \eta~~~~~~~~~~~~~~~~~~~~~~~~~~~~~~~~~~~~~ \nonumber\\
\label{p24001}
\mathrm{s.t.}~~ \frac{B\delta_t}{C^u_i}\sum_{n=1}^{N-1} \hat{R}^{u}_{i}[n]  \geq \eta,\ \ \forall  i,~~~~~~~~~~~~~~~~~~~~~~~~~\\
\frac{B\delta_t}{C^v_j}\sum_{n=2}^{N} \hat{R}^{v}_{j}[n]  \geq \eta,\ \ j=K_3+1,\cdots,V,~~~~~~\\
\label{p24002}
\frac{B\delta_t}{C^v_k} \sum_{n=2}^{N}R^r_{k}[n]  \geq \eta, \ \  k=1,\cdots,K_3,~~~~~\!~~~~~~\\
 \label{p24003}
R^r_{k}[n] \leq \hat{R}^{v}_{k}[n], \ \  k=1,\cdots,K_3, n=2,\cdots, N,\!\!\!\!\\
\label{p24004}
\sum_{m=2}^n R^r_{k}[m] \leq \sum_{m=1}^{n-1} \hat{R}^{u}_{k}[m],~~~~~~~~~~~~~~~~~~~~~~~ \nonumber
\\  k=1,\cdots,K_3, n=2,\cdots,N,\!\!\!\! \\
\eqref{p13008},~~~~~~~~~~~~~~~~~~~~~~~~~~\!~~~~~~~~~~~~~~~~~~~~~ \nonumber
\end{align}
\end{subequations}
where $\hat{R}^{u}_{i}[n]$ and $\hat{R}^{v}_{j}[n]$ are the lower bounds of $R^{u}_{i}[n]$, $R^{v}_{j}[n]$, $\forall i,j,n$, respectively, given in Proposition \ref{theorem01}. Problem (P2.4) can be verified to be convex, which can be efficiently solved by CVX \cite{grant2008cvx}.

With the above two sub-problems solved, (P2.2) can be solved by iteratively optimizing the power and bandwidth allocation and the trajectory with similar steps as in Algorithm \ref{iter1}. Furthermore, the completion time minimization problem in (P2) can be solved via a bisection search over $T$ while solving (P2.2) in each iteration. The details are omitted for brevity.

\section{Initial Trajectory Design}
\label{trajini}

The proposed algorithms for both periodic and one-time operation scenarios require the UAV initial trajectory to be specified, and their converged results via the successive convex optimization and block coordinate descent techniques depend on the UAV trajectory initialization in general. In this section, we propose new trajectory initialization schemes for the periodic and one-time operation scenarios, respectively. Note that due to the additional {\it information-causality constraints} for the one-time operation scenario, these two operation scenarios generally require different trajectory initializations.

Intuitively, the UAV trajectory should be designed so that when the UAV is scheduled to communicate with a particular GU, it should be as close to the GU as possible. One intuitive approach for trajectory initialization is to minimize the UAV traveling time $T_{\mathrm{tr}}$ among different GUs, so that when the given value of $T$ is sufficiently large, the UAV will be able to reach the top of all GUs to enjoy the best communication channel. Given the maximum speed $V_{\mathrm{max}}$, the problem of minimizing the traveling time is thus equivalent to minimizing the total traveling distance. The above approach will be used in the following designs, as detailed later.

\vspace{-0.2cm}
\subsection{Initial Trajectory Design for Periodic Operation}
\label{trajiniup}

In this subsection, we design the UAV initial trajectory for Algorithm \ref{iter1} in the case of periodic operation for any given value of $T$. For notational convenience, let $\mathcal{E}$ denote the set containing all GUs, i.e., $\mathcal{E}\triangleq \mathcal{U}\cup \mathcal{V} =\{ 1, \cdots,U,U+1,\cdots,U+V\}$, where GUs $1,\cdots,U$ correspond to source GUs in $\mathcal{U}$ while GUs $U+1,\cdots,U+V$ correspond to destination GUs in $\mathcal{V}$. The locations of all GUs in $\mathcal{E}$, $\{\textbf{s}_i\}_{i=1}^U$ and $\{\textbf{d}_j\}_{j=1}^V$, are compactly denoted as $\textbf{e}_w$, where $w$ represents the index of the GU in $\mathcal{E}$.

For given $\{\textbf{e}_w\}$, we first consider the problem of minimizing the traveling distance/time for the UAV to visit all GUs by determining their optimal visiting orders, which is essentially the classic TSP \cite{laporte1992traveling}. Although TSP is NP-hard, various algorithms have been proposed to find high-quality approximate solutions within a reasonable computational complexity \cite{laporte1992traveling}. After solving the TSP, we obtain the minimum traveling time required, denoted as $T_{\mathrm{tsp}}$, as well as the permutation order $\hat{\boldsymbol{\pi}}\triangleq [\hat{\pi}(1),\cdots,\hat{\pi}(U+V) ]$, with $\hat{\pi}(w) \in \mathcal{E}$ representing the index of the $w$th GU to be visited. In the following, for any given flight period $T$, the UAV initial trajectory is designed by distinguishing two cases, depending on whether $T$ is no smaller than $T_{\mathrm{tsp}}$, as follows.

$\!\!\!\!\!\!\bullet$ Case 1: $T\geq T_{\mathrm{tsp}}$. In this case, $T$ is sufficiently large so that the UAV is able to reach the top of each GU within each flight period. The remaining time $T-T_{\mathrm{tsp}}$ can be spent by the UAV to hover above the GUs. To obtain an effective method for determining the hovering time allocation $\tilde{T}_w$ among the GUs, let $\tilde{T}'_w$ denote the time required for the UAV to satisfy the average rate requirement for GU $w$, by assuming that the UAV only communicates with it when hovering on its top. We thus have
\vspace{-0.2cm}
\begin{align}
\widetilde{T}'_w= \left\{
 \begin{array}{ll}
    \frac{T\bar{R}_w^u}{B \log_2\left(1+\frac{P_w^u\gamma_0}{H^2} \right)}, & w=1,\cdots,U, \\
    \frac{T \bar{R}_{w-U}^v}{B \log_2 \left(1+\frac{P^v \gamma_0}{H^2} \right) }, & w=U+1,\cdots,U+V.
  \end{array}
\right.
\end{align}

Then the total hovering time $T- T_{\mathrm{tsp}}$ can be proportionally divided among the GUs as
\begin{align}
\label{hover}
\widetilde{T}_w=\frac{\widetilde{T}'_w(T-T_{\mathrm{tsp}})}{\sum_{y=1}^{U+V}\widetilde{T}'_y}, \ \ w\in \mathcal{E}.
\end{align}

Following the visiting order $\hat{\boldsymbol{\pi}}$ and the hovering time allocation in \eqref{hover} for each GU, the initial trajectory $\mathcal{Q}^0$ in the case of $T \geq T_{\mathrm{tsp}}$ can be constructed accordingly.

\begin{figure}
\vspace{-0.4cm}
\centering
\includegraphics[width=4.5in]{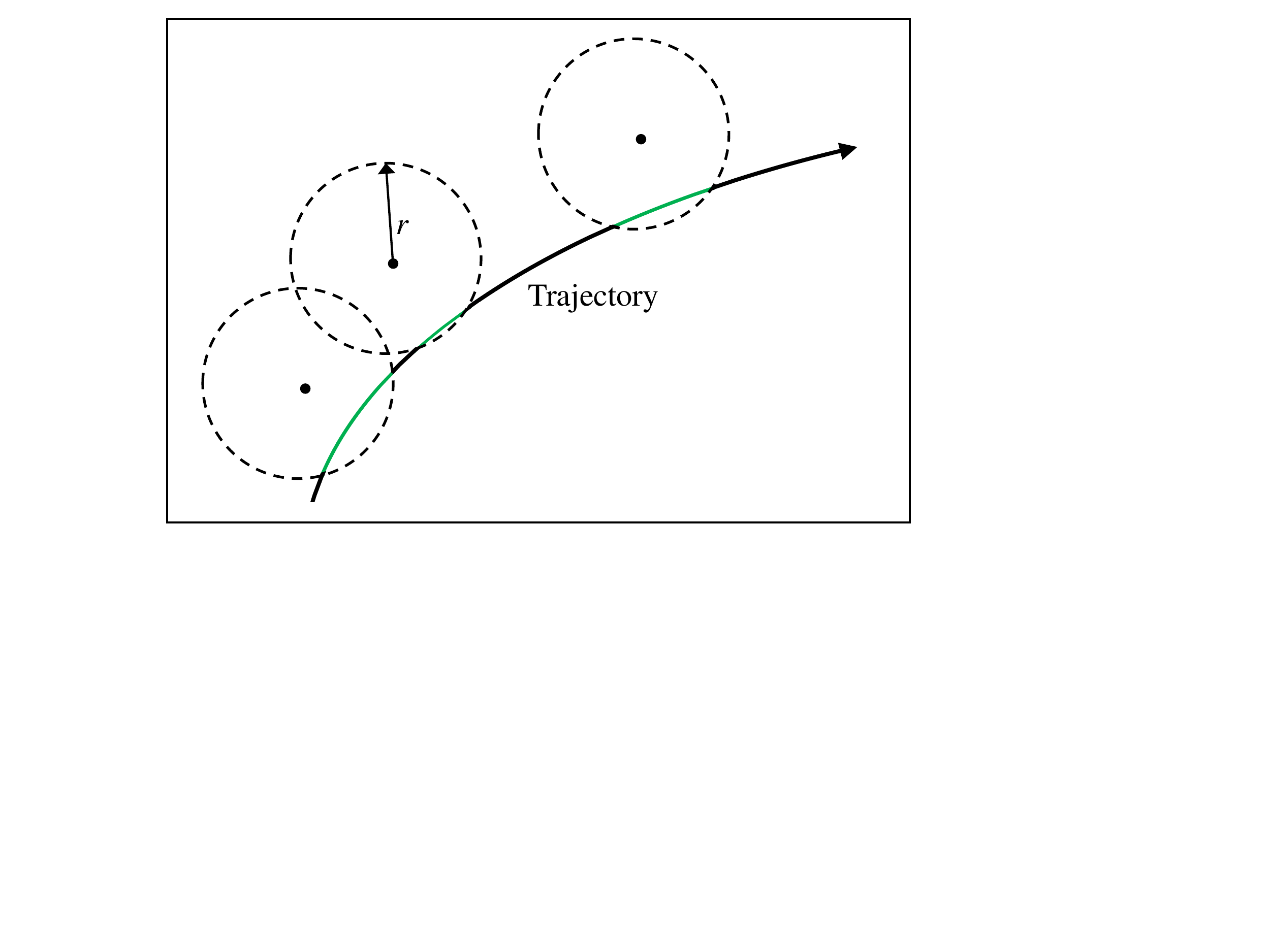}
\vspace{-4.3cm}
\caption{\label{tspn}An illustration of the disk-shaped region for the UAV to reach.}
\vspace{-0.4cm}
\end{figure}

$\!\!\!\!\!\!\bullet$ Case 2: $T< T_{\mathrm{tsp}}$. In this case, the given time $T$ is insufficient for the UAV to reach the top of all GUs. To design a feasible initial trajectory, we first specify a disk-shaped region for each GU in $\mathcal{E}$, which is centered at the corresponding GU with radius $r$. As illustrated in Fig. \ref{tspn}, the main idea is to minimize the UAV traveling distance by properly designing the UAV trajectory and radius $r$, so that the UAV is able to reach each disk region. The problem can be formulated as
\vspace{-0.2cm}
\begin{subequations}
\begin{align}
\mathrm{(P3)}~~ \min_{r, \textbf{q}(t),T_{\mathrm{tr}} } ~~T_{\mathrm{tr}}\nonumber~~~~~~~~~~~~~~~~~~~~~~~~~\\
\label{p3001}
\mathrm{s.t.}~~\min_{0\leq t \leq T_{\mathrm{tr}}} ||\textbf{q}(t)-\textbf{e}_w||\leq r, \ \ \forall w \in \mathcal{E}, \!\!\!\!\\
\label{p3002}
||\dot{\textbf{q}}(t)|| \leq V_{\mathrm{max}}, \ \ \forall 0\leq t \leq T_{\mathrm{tr}},\!\!\!~~~\!\!\\
\label{p3003}
\textbf{q}(0)=\textbf{q}(T_{\mathrm{tr}}),~~~~~~~~~~~\!~~~~\!~~~~~~
\end{align}
\end{subequations}
where constraints \eqref{p3001} ensure that for each GU in $\mathcal{E}$, there exists at least one time instant $t$ such that the distance between the UAV and the GU is no larger than $r$. This guarantees that all disks are traversed by the UAV.

For any given radius $r$, denote the optimal value of (P3) as $T^*_{\mathrm{tr}}(r)$. It is not difficult to see that $T^*_{\mathrm{tr}}(r)$ is non-increasing with $r$. Thus, the optimal solution to (P3) can be obtained by solving the corresponding problem with fixed $r$, and then applying a bisection search to find the optimal radius $r$. In the following, we focus on solving (P3) with any given radius $r$.
\vspace{-0.2cm}
\begin{Lemma}
\label{lemmatraj}
The optimal trajectory to (P3) should only contain connected line segments.
\end{Lemma}
\vspace{-0.2cm}
\begin{IEEEproof}
Similar to the proof of Theorem 1 in \cite{8255824}, Lemma \ref{lemmatraj} can be shown by contradiction. Suppose on the contrary that at the optimal solution $(\textbf{q}^*(t), T^*_{\mathrm{tr}})$, there exists at least one curved portion along the trajectory. Then we can always construct an alternative trajectory $\textbf{q}'(t)$ composed of line segments only, that achieves less traveling time $T'_{\mathrm{tr}}< T^*_{\mathrm{tr}}$. To this end, it is first noted that at the optimal solution to (P3), the UAV should always travel with the maximum speed $V_{\mathrm{max}}$, i.e., the constraint \eqref{p3002} should be satisfied with equality.

For each node with $w\in \mathcal{E}$, denote by $E_w$ as the earliest time instance when the UAV reaches its disk region with trajectory $\textbf{q}^*(t)$, i.e., $E_w$$\triangleq $$\min\{0 \! \leq \! t \! \leq \! T^*_{\mathrm{tr}}~|~|| \textbf{q}^*(t)-\textbf{e}_w ||\!\leq \! r \} $. Then the trajectory $\textbf{q}^*(t)$ of the UAV can be partitioned into $U+V+1$ portions, with the $w$th portion specified by time interval $[E_{w-1},E_w]$, $w$$=$$1,\cdots,U+V+1$, where $E_0\triangleq 0$, $E_{U+V+1}\triangleq T^*_{\mathrm{tr}}$. For the $w$th time interval, we may replace the original trajectory portion with a line segment directly connecting $\textbf{q}'_{w-1}\triangleq\textbf{q}^*(E_{w-1})$ and $\textbf{q}'_w\triangleq\textbf{q}^*(E_{w})$. Obviously, this replacement not only ensures the feasibility~of \eqref{p3001}, but also reduces the traveling distance for the UAV. Therefore, if the optimal solution trajectory $\textbf{q}^*(t)$ contains a curved portion, we are always able to construct an alternative trajectory by sequentially connecting $\textbf{q}'_0, \cdots, \textbf{q}'_{U+V+1}$ that achieves $T'_{\mathrm{tr}}$$< $$T^*_{\mathrm{tr}}$. Thus, any trajectory with curved portion cannot be the optimal trajectory to (P3). This completes the proof.
\end{IEEEproof}

With Lemma \ref{lemmatraj}, for any given $r$, problem (P3) is recast to optimizing a set of waypoints inside the disks, which are the starting and ending points of the line segments, and finding the optimal permutation order $\boldsymbol{\pi}\triangleq [\pi(1),\cdots,\pi(U+V)  ]$ to visit these waypoints. Let the waypoint inside the disk associated with GU $w$ be denoted as $\textbf{\emph{g}}_{w}\in \mathbb{R}^{2 \times 1}$, $w=1,\cdots,U+V$. The traveling time required can be expressed as
\vspace{-0.2cm}
\begin{align}
T_{\mathrm{tr}}(\{\textbf{g}_w\},\boldsymbol{\pi})=~~~~~~~~~~~~~~~~~~~~~~~~~~~~~~~~~~~~~~~~~~~~\nonumber\\
\frac{\sum_{w=1}^{{U+V-1}}||\emph{\textbf{g}}_{\pi({w+1})}-\emph{\textbf{g}}_{\pi(w)}||+||\emph{\textbf{g}}_{\pi({U+V})}-\textbf{\emph{g}}_{\pi(1)}||}{V_{\mathrm{max}}}.\!\!\!
\end{align}
As a result, problem (P3) reduces to
\vspace{-0.2cm}
\begin{align}
\mathrm{(P3.1)}~~ \min_{\{\textbf{\emph{g}}_w\},\boldsymbol{\pi} } ~~T_{\mathrm{tr}}(\{\textbf{g}_w\},\boldsymbol{\pi} ) ~~~~~~~~~~ \nonumber \\
\mathrm{s.t.}~~~|| \textbf{\emph{g}}_{w}-\textbf{e}_{w}||\leq r, \ \ w\in \mathcal{E}.~~~\!~~
\end{align}
This is reminiscent of the classic Traveling Salesman Problem with Neighborhoods (TSPN) \cite{arkin1994approximation}, which is a generalization of TSP and also known to be NP-hard \cite{safra2006complexity}.

In the following, we propose an efficient approach to find~a suboptimal solution to (P3.1). The key idea is to let the~UAV visit each disk region based on the order $\hat{\boldsymbol{\pi}}$ obtained~by~the TSP algorithm (by ignoring the neighborhoods), i.e., $\boldsymbol{\pi}=\hat{\boldsymbol{\pi}}$, and then find the optimal waypoints inside all disks by solving a similar convex optimization problem as in \cite{8255824}. With the visiting order obtained, (P3.1) is recast to a convex optimization problem, which can be efficiently solved via standard convex optimization techniques.

It is worth noting that TSPN has been extensively studied in the literature (e.g., \cite{4288144}, \cite{yuan2010racetrack} and \cite{6197191}). Based on the permutation obtained by the TSP algorithm, the authors in \cite{4288144} adopted three evolutionary algorithms to find the shortest path with disjoint disks only. In \cite{6197191}, the authors proposed a combine-skip-substitute (CSS) scheme based on TSP, which is applicable to both joint or disjoint disks. However, there is no guarantee that the optimal waypoints can be found even with given visiting order. In contrast, by applying convex optimization in this work, the optimal waypoints are guaranteed with the given visiting order.

Combining the above Case 1 and Case 2, the design of the initial trajectory for Algorithm \ref{iter1} in the periodic operation case with given flight period $T$ is summarized in Algorithm \ref{algo3}.
\vspace{-0.2cm}
\begin{algorithm}
\caption{Trajectory initialization for periodic operation with given flight period $T$.\label{algo3}}
1:~~Solve the TSP to obtain the traveling time $T_{\mathrm{tsp}}$ and visiting  \\
\hspace*{0.5cm}order $\hat{\boldsymbol{\pi}}$; let tolerance $\epsilon_1> 0$.\\
2:~~\textbf{if} $T\geq T_{\mathrm{tsp}}$\\
3:~~~~~Construct the initial trajectory according to Case 1.\\
4:~~\textbf{else}\\
5:~~~~~Let $r_1=0$, $r_2$ be sufficiently large.\\
6:~~~~~~~\textbf{repeat}\\
7:~~~~~~~\!\!~~Update $r=(r_1+r_2)/2$.\\
8:~~~~Based on visiting order $\hat{\boldsymbol{\pi}}$, obtain the traveling
 \hspace*{1.15cm}  time $T_{\mathrm{tr}}$ by solving\ (P3.1). \\
9:~~~~~~~~If $T_{\mathrm{tr}}> T$, let $r_1=r$. Else, let $r_2=r$.\\
10:~~~~~~\!\textbf{until} $(r_2-r_1)\leq\epsilon_1$.\\
11:~~\!~~Construct the initial trajectory according to Case 2.\\
12:~\!~\textbf{end}
\end{algorithm}
\subsection{Initial Trajectory Design for One-Time Operation}
\label{inionetime}

In this subsection, we propose an efficient trajectory initialization for the one-time operation. Different from periodic operation, one-time operation is subject to the additional \emph{information-causality constraints} \eqref{infcau} for the data relaying service, which needs to explicitly take into account the visiting order of the corresponding GUs, i.e., before approaching the destination GU for information forwarding, the UAV should first fly closer to the corresponding source GU to collect data. In this case, the TSP-based trajectory initialization usually leads to poor performance (as verified  by simulations in Section \ref{simulation}) since it ignores the visiting order for such GUs. In the following, we propose a new initial trajectory design by taking into account such precedence consideration.

With the above precedence consideration, minimizing the traveling distance of visiting all GUs is reminiscent of the classic PDP, which is also known as dial-a-ride problem (DARP) \cite{parragh2008survey}. A brief description of PDP and its variations are given in Appendix \ref{over_pdp}. Note that the corresponding precedence constraints only apply for GUs in Group 3 while such constraints are irrelevant for GUs in Group 1 and Group~2. Therefore, the problem of minimizing the traveling distance~to visit all GUs in one-time operation is a hybrid of TSP and~PDP. However, for ease of presentation, we will mainly relate it to PDP.

In set $\mathcal{E}$, recall that GUs $k$ and $U+k$, $k=1,\cdots,K_3$, form a pair of source-destination GUs for data relaying. For GUs $a,b \in \mathcal{E}$, $a\neq b$, we define the traveling cost between $a$ and $b$ as $c_{a,b}\triangleq||\textbf{e}_{a}-\textbf{e}_{b}||$ and the associated traveling time as $t_{a,b}\triangleq||\textbf{e}_{a}-\textbf{e}_{b}||/V_{\mathrm{max}}$. As discussed in \cite{8255824}, since the considered problem does not require the UAV to return to the initial location, we may introduce a dummy GU 0, whose distances to all other GUs in $\mathcal{E}$ are 0, i.e., $c_{a,0}=c_{0,b}=t_{a,0}=t_{b,0}=0$, $a,b \in \mathcal{E}$. As a result, a new GU set can be defined as $\widetilde{\mathcal{E}}\triangleq \mathcal{E} \cup \{0\}$. We then define a binary variable $x_{a,b}$, $a, b \in \widetilde{\mathcal{E}}$, $a\neq b$, which chooses 1 if the edge connecting $a$ and $b$ is traversed by the UAV and 0 otherwise. Then the traveling time can be expressed as $T_{\mathrm{tr}}=\left(\sum_{a \in \widetilde{\mathcal{E}}} \sum_{b\in \widetilde{\mathcal{E}}, b\neq a}c_{a,b}x_{a,b}\right)/V_{\mathrm{max}}$. Further denoting by $T_w$ the time when the UAV reaches the GU $w$, $w \in \mathcal{E}$, and $T_0\triangleq0$ the starting time of the UAV from GU 0, the problem can be formulated as
\vspace{-0.1cm}
\begin{subequations}
\begin{align}
\mathrm{(P4)} \min_{\substack{\{T_w\}\\x_{a,b}, a,b\in \widetilde{\mathcal{E}}}}~T_{\mathrm{tr}} \nonumber~~~~~ ~~~~~~~~~~~~~~~~~~~~~~~\!~~~\\
\label{p4001}
\mathrm{s.t.}~\sum_{a\in \widetilde{\mathcal{E}},a\neq b}  x_{a,b}=1,\ \ \forall b\in \widetilde{\mathcal{E}} , ~~~~~~~~~~~~~~\\
\label{p4002}
\sum_{b\in \tilde{\mathcal{\mathcal{E}}},a\neq b } x_{a,b}=1,\ \ \forall a\in \widetilde{\mathcal{E}},~~\!~~~~~~\!~~~~~~~\!\\
\label{p4003}
T_0 \leq T_w, \ \ w\in \mathcal{E} ,~~~~~~~~~~~~~~~~~~~~~~~ \\
\label{p4004}
 (T_a+t_{a,b})x_{a,b} \leq T_b, \ \ \forall a,b \in \widetilde{\mathcal{E}},a \neq b, \!\!\!\\
\label{p4005}
 T_k \leq T_{U+k}, \ \ k=1,\cdots,K_3,~~~~~~\!~~~~\\
\label{p4006}
x_{a,b}=\{0,1 \},  \ \ \forall a,b \in \tilde{\mathcal{E}}, a \neq b,~~\!~~~\!~~
\end{align}
\end{subequations}
where constraints \eqref{p4001} and \eqref{p4002} guarantee that each GU in $\tilde{\mathcal{\mathcal{E}}}$ is visited exactly once, constraint \eqref{p4003} ensures that the dummy GU 0 is visited first, \eqref{p4004} ensures the consistency of time and \eqref{p4005} corresponds to the precedence constraints that the source GU $k$ is visited before the destination GU $U+k$. Problem (P4) is a mixed-integer optimization problem, which can be solved via CPLEX CP optimizer \cite{WinNT}. It should be noted that without precedence constraints \eqref{p4005}, the dummy GU 0 and its associated constraint \eqref{p4003}, problem (P4) reduces to a standard TSP, which can also be efficiently solved via CPLEX CP optimizer \cite{WinNT}. After solving (P4), the two edges associated with the dummy GU 0 are removed so as to obtain the minimum time required $T_{\mathrm{pdp}}$ for the UAV to visit all GUs, as well as the permutation $\hat{\boldsymbol{\phi}}\triangleq [\hat{\phi}(1),\cdots,\hat{\phi}(U+V)]$, with $\hat{\phi}(w)\in \mathcal{E}$ representing the index of the GU to be visited. Similarly as in Section \ref{trajiniup}, the initial trajectory for the one-time operation will be designed by distinguishing whether $T$ is no smaller than the obtained $T_{\mathrm{pdp}}$, as follows.

$\!\!\!\!\!\!\bullet$ Case 1: $T\geq T_{\mathrm{pdp}}$. In this case, the UAV is able to reach all the GUs with time $T_{\mathrm{pdp}}$ and the remaining time $T-T_{\mathrm{pdp}}$ can be proportionally divided among the GUs similarly as in~\eqref{hover}.

$\!\!\!\!\!\!\bullet$ Case 2: $T< T_{\mathrm{pdp}}$. In this case, the given time $T$ is insufficient for the UAV to reach all GUs. Similar to Section \ref{trajiniup}, a disk-shaped region is specified for each GU with radius $r$ such that the UAV initial trajectory is designed to ensure that it reaches the disk region of each GU. Due to the {\it information-causality constraints}, the resulted precedence constraints should also be imposed in this case to guarantee that the UAV visits the disk of the source GU $k$ before that of the destination GU $U+k$, $k=1,\cdots,K_3$. Under the new precedence constraints over certain disks, the problem here is to design a trajectory traversing all disks with the minimum traveling distance, which we refer to as Pickup-and-Delivery Problem with Neighborhoods (PDPN).

Similar to (P3), as the traveling time $T_{\mathrm{tr}}$ is non-increasing with the radius $r$, the problem of minimizing the traveling distance under precedence constraints can also be solved via a bisection search over $r$. By following the similar proof in Lemma \ref{lemmatraj}, for a fixed radius $r$, the problem can be further recast to optimizing a set of waypoints inside disks and finding the optimal permutation order $\boldsymbol{\phi}\triangleq[\phi(1),\cdots,\phi(U+V)] $ to visit these disks. With the precedence constraints over disks involved, an efficient suboptimal solution can be obtained by letting the UAV visit each disk region following the order $\hat{\boldsymbol{\phi}}$ (without considering neighborhoods) obtained by solving (P4), i.e., $\boldsymbol{\phi}=\hat{\boldsymbol{\phi}}$, and applying convex optimization technique to find the optimal waypoints inside disks, similarly as for (P3.1). Then the initial trajectory can be constructed~accordingly.

As a summary, for any given mission time $T$, the initial trajectory for the one-time operation scenario can be constructed with similar steps as in Algorithm \ref{algo3}. The details are omitted for brevity.

\section{Simulation Results}
\label{simulation}

\begin{table}[]
\centering
\caption{Parameter values for numerical simulations.}
\label{tab01}
\begin{tabular}{|l|l|}
\hline
UAV altitude                   & $H=50$ m                                 \\ \hline
Maximum UAV speed              & $V_{\mathrm{max}}=50$ m/s            \\ \hline
Transmit power of source GUs & $P^u_{1}=P^u_{2}=P^u_{3}=0.01$ W \\ \hline
Transmit power of UAV & $P^v=0.01$ W                              \\ \hline
Bandwidth                      & $B=10$ MHz                                \\ \hline
Channel power at reference distance $\tilde{d}_0$=1 m   &            $\lambda_0=-50~\mathrm{dB}$  \\ \hline
Noise power spectrum density    & $N_0=-169$ dBm/Hz                        \\ \hline
\end{tabular}
\vspace{-0.5cm}
\end{table}

In this section, numerical results are provided to evaluate the performance of our proposed designs. We consider a system with $K=6$ GUs, three source GUs and three destination GUs, i.e., $U=V=3$, which are randomly and uniformly distributed in a square area of side length equal to 6000 m. The following results are based on one realization of GUs' locations shown in Fig. \ref{periodc2}. We assume that all GUs have equal rate requirement, i.e., $\bar{R}\triangleq\bar{R}_i^u=\bar{R}_j^v$, $C\triangleq C_i^u=C_j^v$, $\forall i,j$. Unless otherwise stated, the parameter values are given in Table \ref{tab01}. As a benchmark comparison with our proposed trajectory initializations in Section \ref{trajini}, the circular trajectory initialization in \cite{8254949} is considered.

\subsection{Periodic Operation}

\begin{figure*}
\vspace{-3.8cm}
\begin{minipage}{0.6\linewidth}
\includegraphics[width=3.8in]{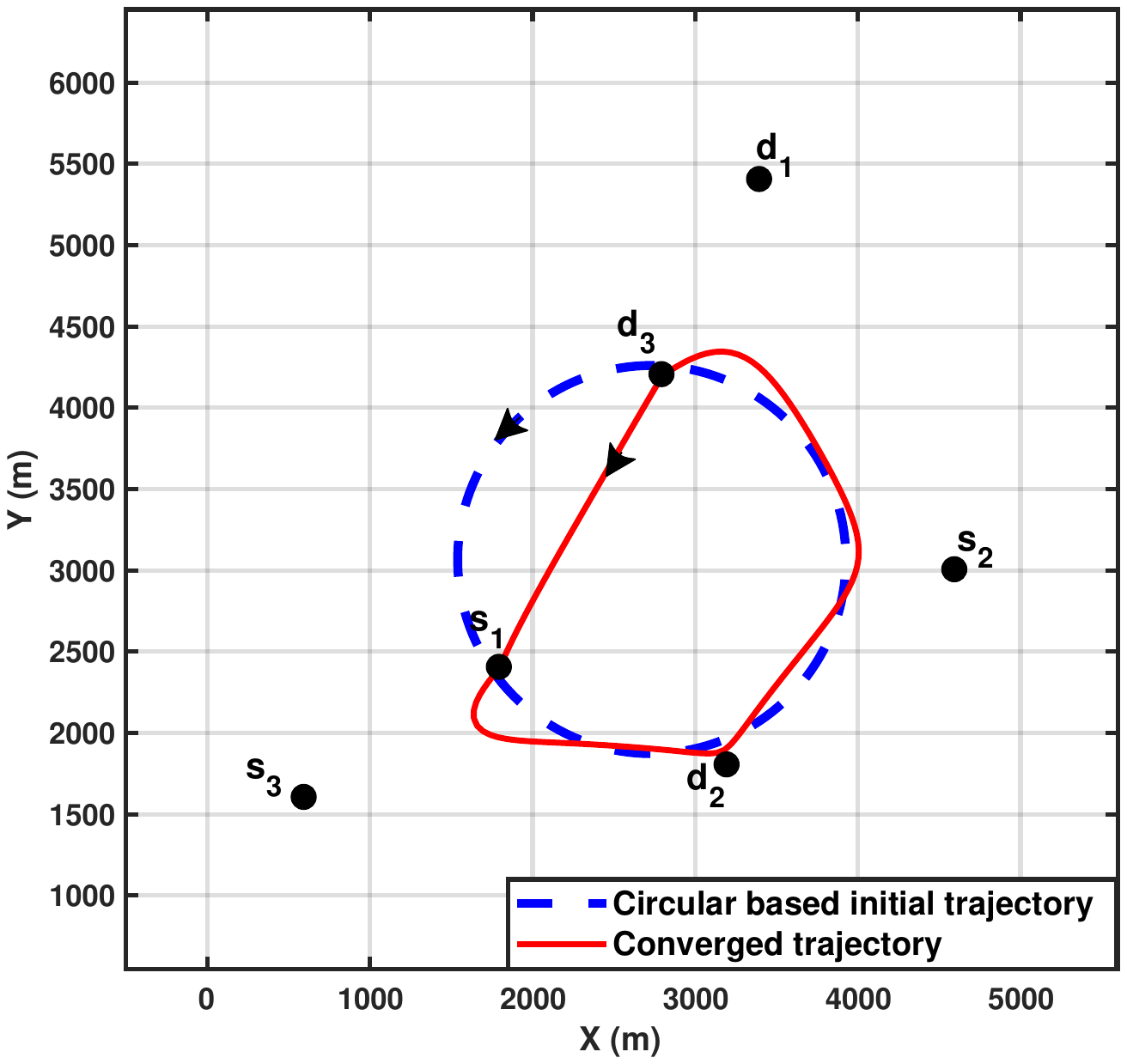}
\vspace{-3.8cm}
\subcaption{\label{periodc2}Circular based initialization.~~~~~~  }
\end{minipage}
\hspace*{-2.6cm}\begin{minipage}{0.6\linewidth}
\includegraphics[width=3.8in]{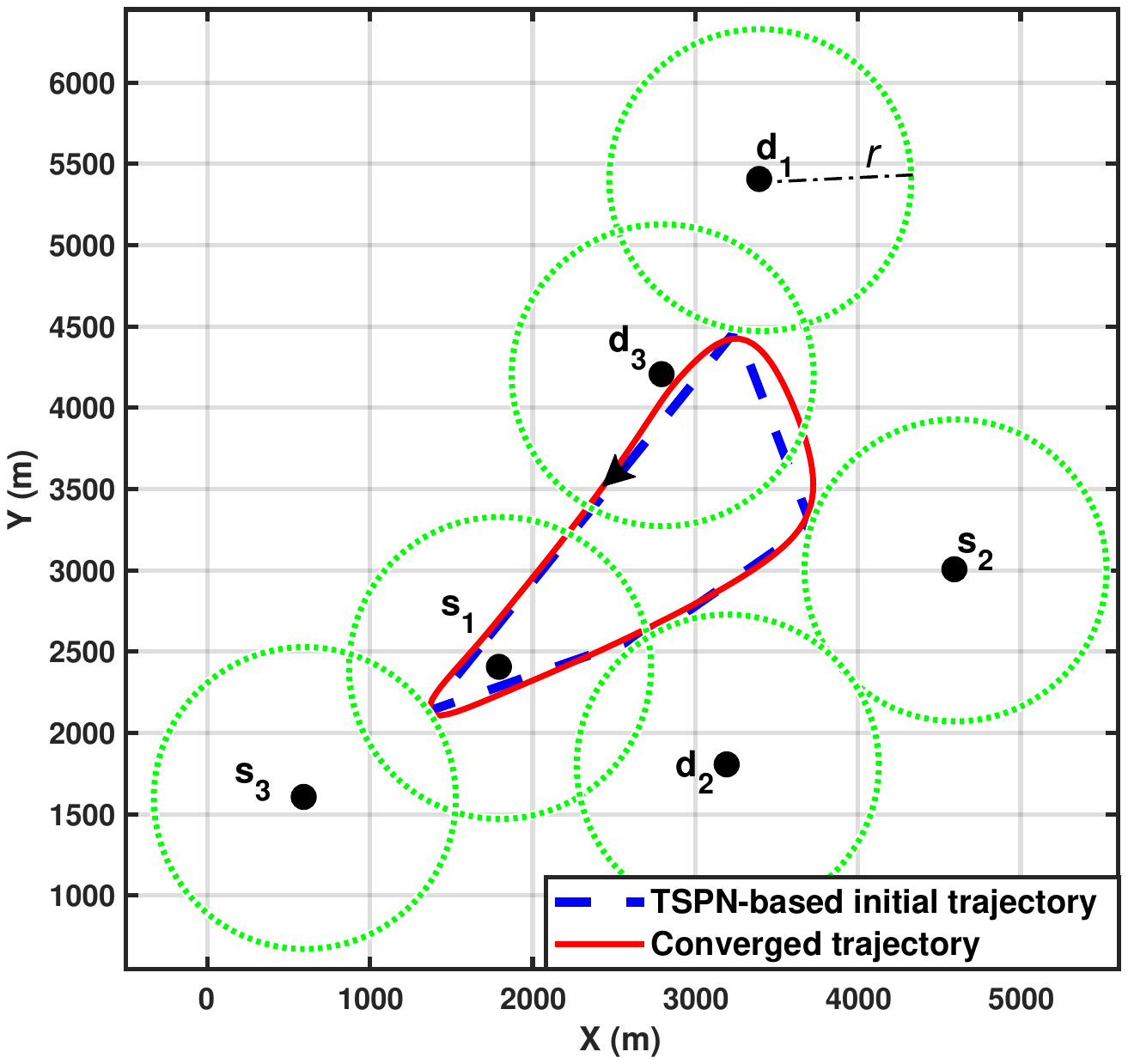}
\vspace{-3.8cm}
\subcaption{\label{periodt2}TSPN-based initialization ($r=928$ m).}
\end{minipage}
\vspace{-0.2cm}
\caption{\label{period2}UAV trajectories with different initializations under average rate requirement $\bar{R}=2$ Mbps for periodic operation.}
\vspace{-0.6cm}
\end{figure*}
\begin{figure*}
\vspace{-2.8cm}
\begin{minipage}{0.6\linewidth}
\includegraphics[width=3.8in]{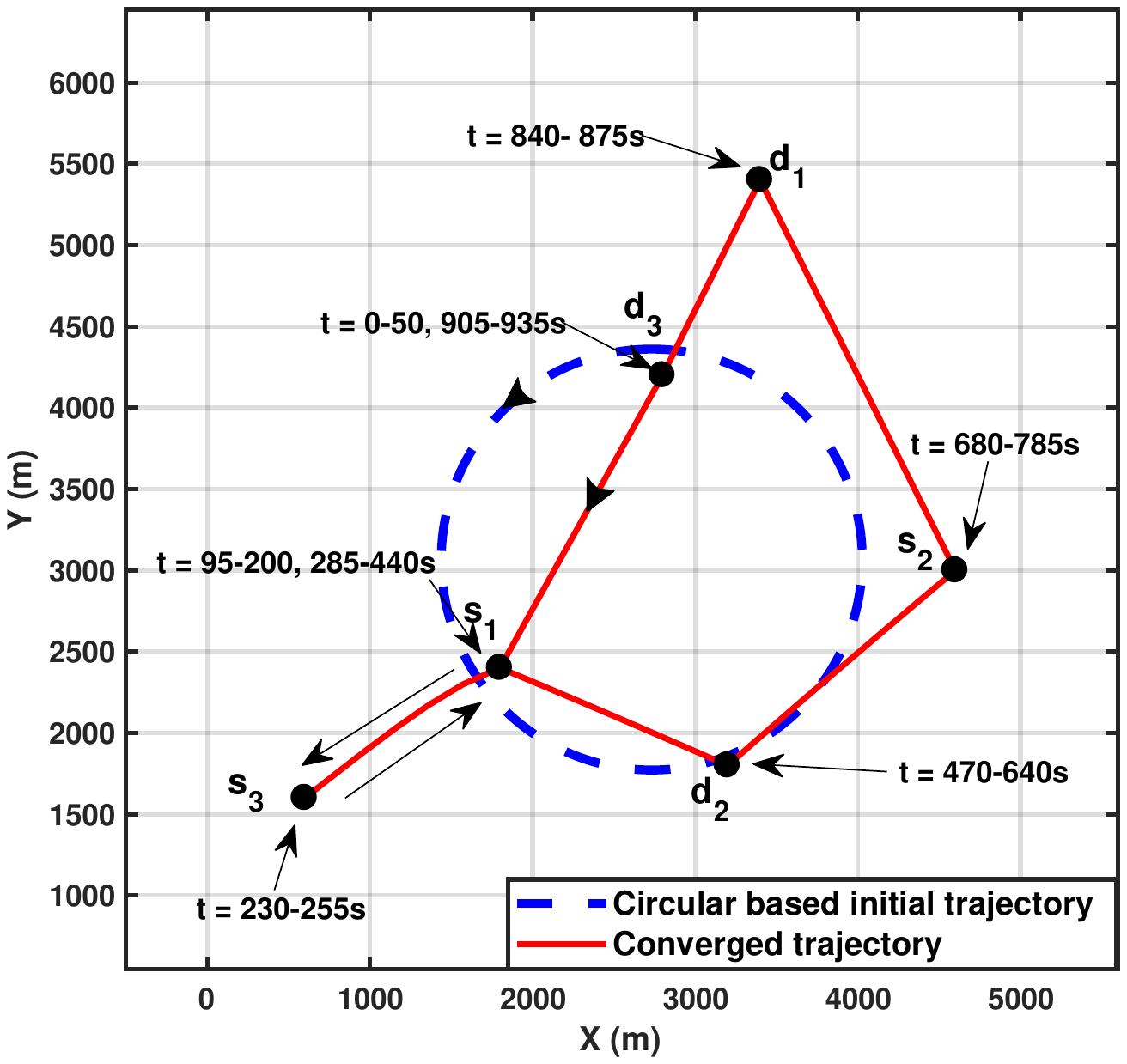}
\vspace{-3.8cm}
\subcaption{\label{periodc5_5}Circular based initialization.~  }
\end{minipage}
\hspace*{-2.6cm}\begin{minipage}{0.6\linewidth}
\includegraphics[width=3.8in]{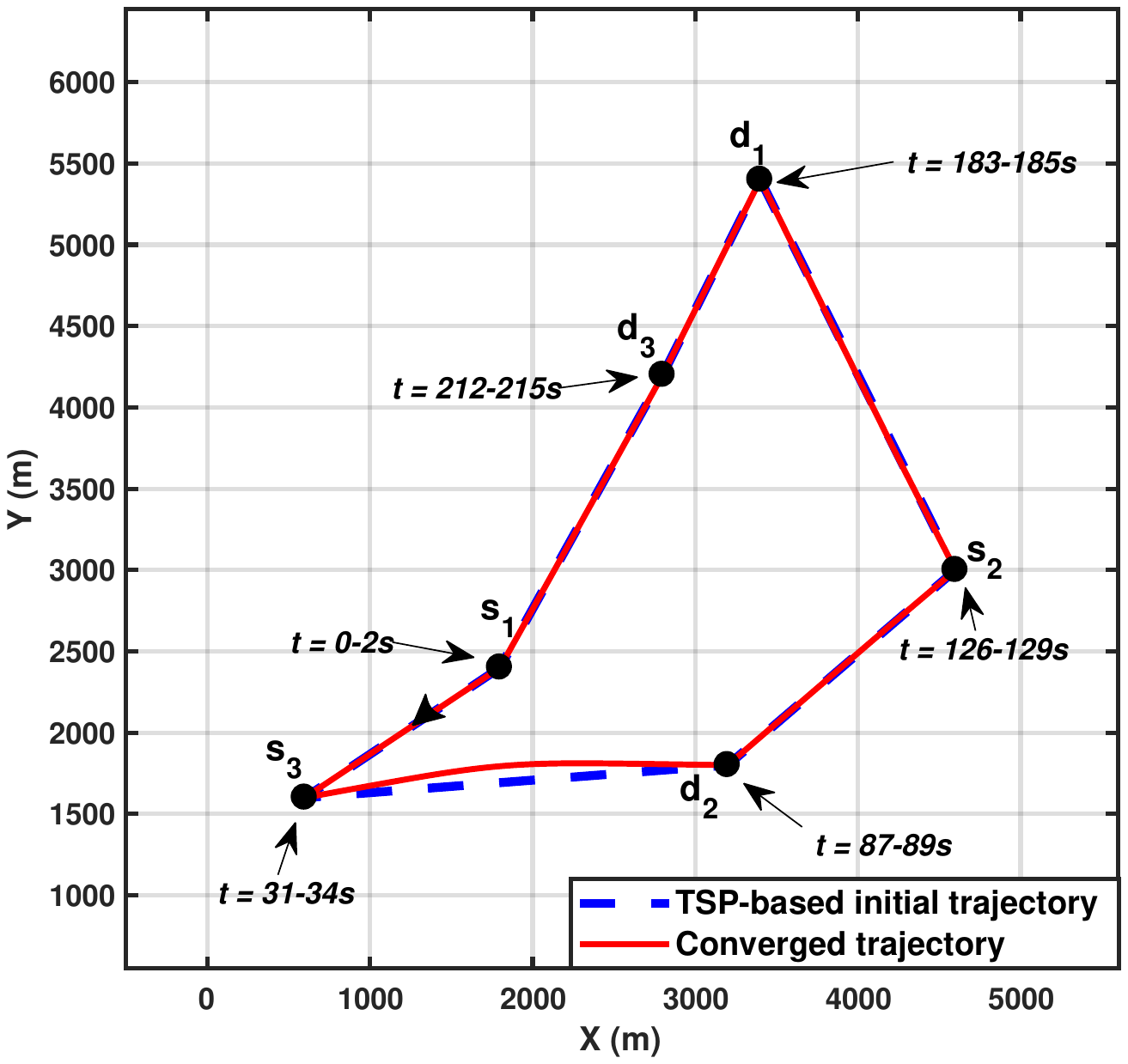}
\vspace{-3.8cm}
\subcaption{\label{periodt5_5}TSP-based initialization.~}
\end{minipage}
\vspace{-0.2cm}
\caption{\label{period5_5}UAV trajectories with different initializations under average rate requirement $\bar{R}=5.5$ Mbps for periodic operation.}
\vspace{-0.5cm}
\end{figure*}

In this subsection, we focus on the periodic operation studied in Section \ref{period} and compare the performances between the circular based trajectory initialization and the TSP/TSPN-based trajectory initialization proposed in Section \ref{trajiniup}. The corresponding TSP in Section \ref{trajiniup} is solved by CPLEX CP Optimizer \cite{WinNT} via a transformation of (P4) as discussed previously. By solving the TSP with given GUs' locations in Fig. \ref{periodc2}, the minimum time required for the UAV to visit all GUs can be obtained as $T_{\mathrm{tsp}}=239$ s.

Under the average rate requirement of $\bar{R}=2$ Mbps, the obtained UAV trajectories for circular based initialization~and TSPN-based initialization are respectively shown in Figs. \ref{periodc2} and \ref{periodt2}, with the corresponding flight periods obtained~as 150 s and 139 s, respectively. It is observed that for both schemes, the UAV does not have to fly to the top of all GUs for communications. This is expected, since with relatively low rate requirement of $\bar{R}=2$ Mbps, the communication links are sufficiently good even when the UAV has some moderate distance from the GUs. Furthermore, it is observed that the proposed TSPN-based initialization scheme in general results in different trajectories from the circular based initialization, and it requires smaller flight duration in each period (139 s versus 150 s) under the same rate requirement.

As the average rate requirement increases to $\bar{R}=5.5$ Mbps, the obtained trajectories are shown in Fig. \ref{period5_5}. By comparing with Fig. \ref{period2}, it is observed that in this case, the UAV needs to fly to the top of each GU to enjoy the best communication link quality, which is expected due to the high rate requirement. Furthermore, it is observed that the circular based initialization results in a visiting order that is different from that with the TSP-based initialization. Specifically, with circular based initialization, the UAV will start with flying from $\textbf{d}_3$, and after visiting $\textbf{s}_1$ and $\textbf{s}_3$, it will revisit $\textbf{s}_1$, rather than  directly fly towards $\textbf{d}_2$ as with the TSP-based initialization shown in Fig. \ref{periodt5_5}. This is obviously undesirable since it unnecessarily increases the traveling time in the converged trajectory as compared to the TSP-based initialization. Furthermore, the minimum flight period with the proposed TSP-based initialization is 257 s, which is significantly less than that by the benchmark circular based initialization (935 s). The reason is that with the circular based initial trajectory, at the first iteration of Algorithm \ref{iter1}, a large portion of the power and bandwidth are allocated to GUs far away from the UAV to satisfy various average rate requirements. This becomes a bottleneck for maximizing the minimum ratio $\eta$ and thus results in different hovering time of the UAV above GUs in the converged trajectory. Whereas with equal hovering time above each GU in the TSP-based initial trajectory, power and bandwidth are more efficiently allocated to the GUs that the UAV is hovering above and thus more time is saved.

\begin{figure}
\vspace{-4cm}
\hspace{-1cm}\includegraphics[width=4.5in]{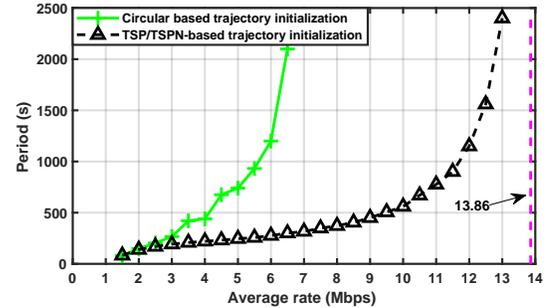}
\vspace{-7.3cm}
\caption{\label{periodcom}Minimum flight period versus average rate requirement for periodic operation. }
\vspace{-0.6cm}
\end{figure}

\begin{figure*}[!htb]
\vspace{-3.5cm}
\minipage{0.32\textwidth}
  \hspace*{-1.9cm}\includegraphics[width=1.65\linewidth]{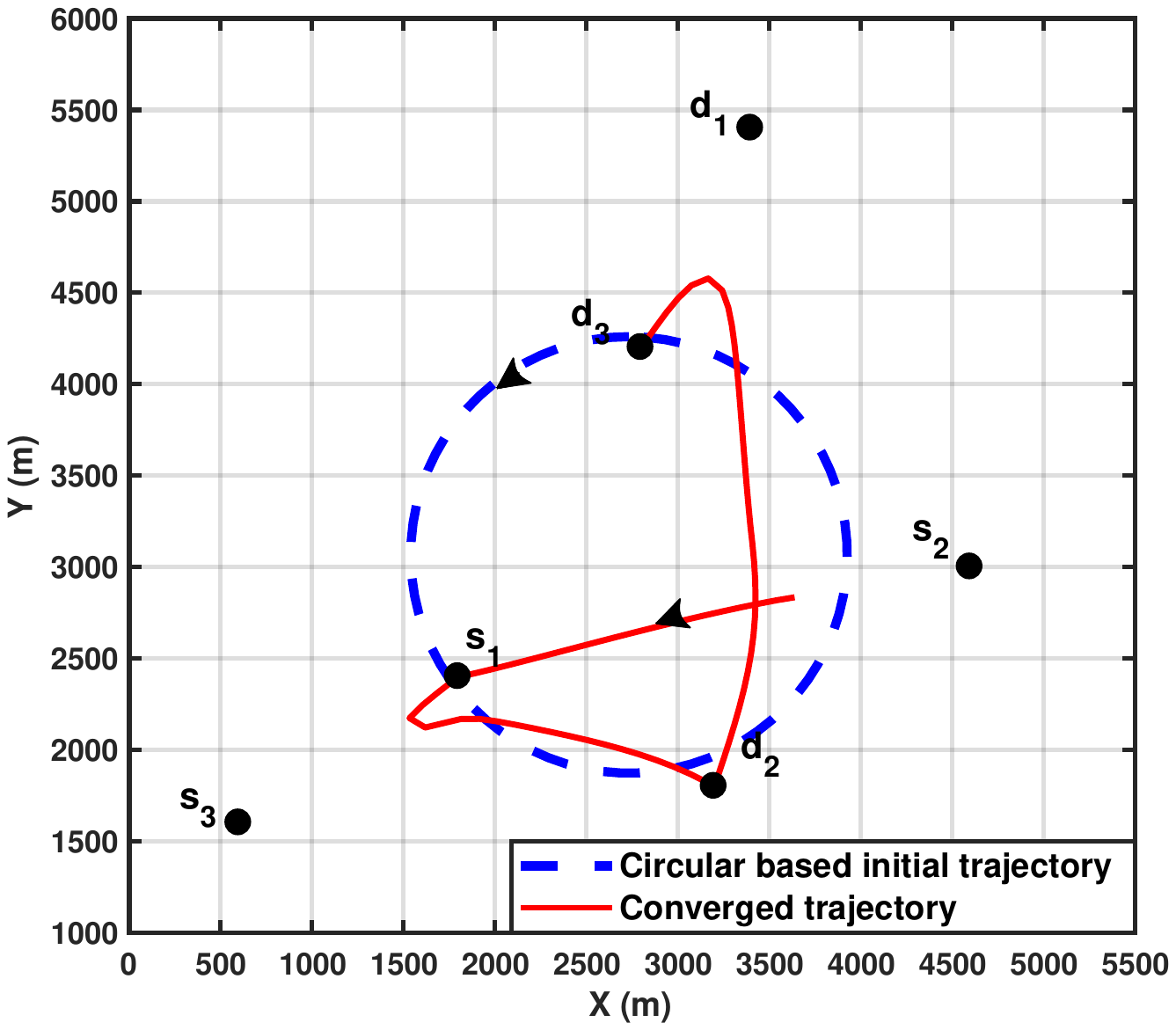}
\vspace{-4.7cm}
  \subcaption{\label{300cir}Circular based initialization.}
\endminipage\hfill
\minipage{0.32\textwidth}
\hspace*{-1.8cm}\includegraphics[width=1.65\linewidth]{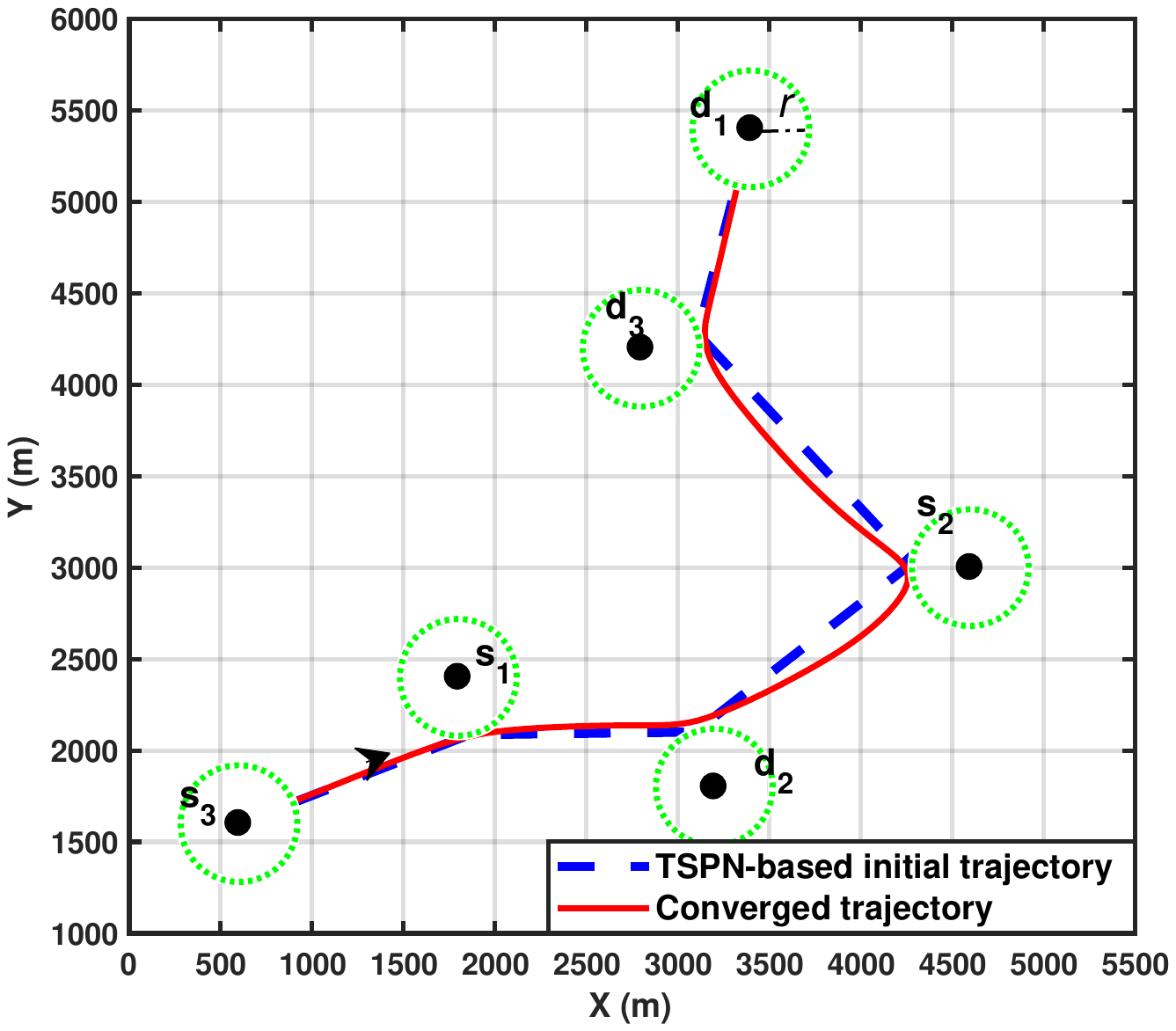}
\vspace{-4.7cm}
  \subcaption{\label{300t}TSPN-based initialization ($r=319$ m).}
\endminipage\hfill
\minipage{0.32\textwidth}%
 \hspace*{-1.7cm} \includegraphics[width=1.65\linewidth]{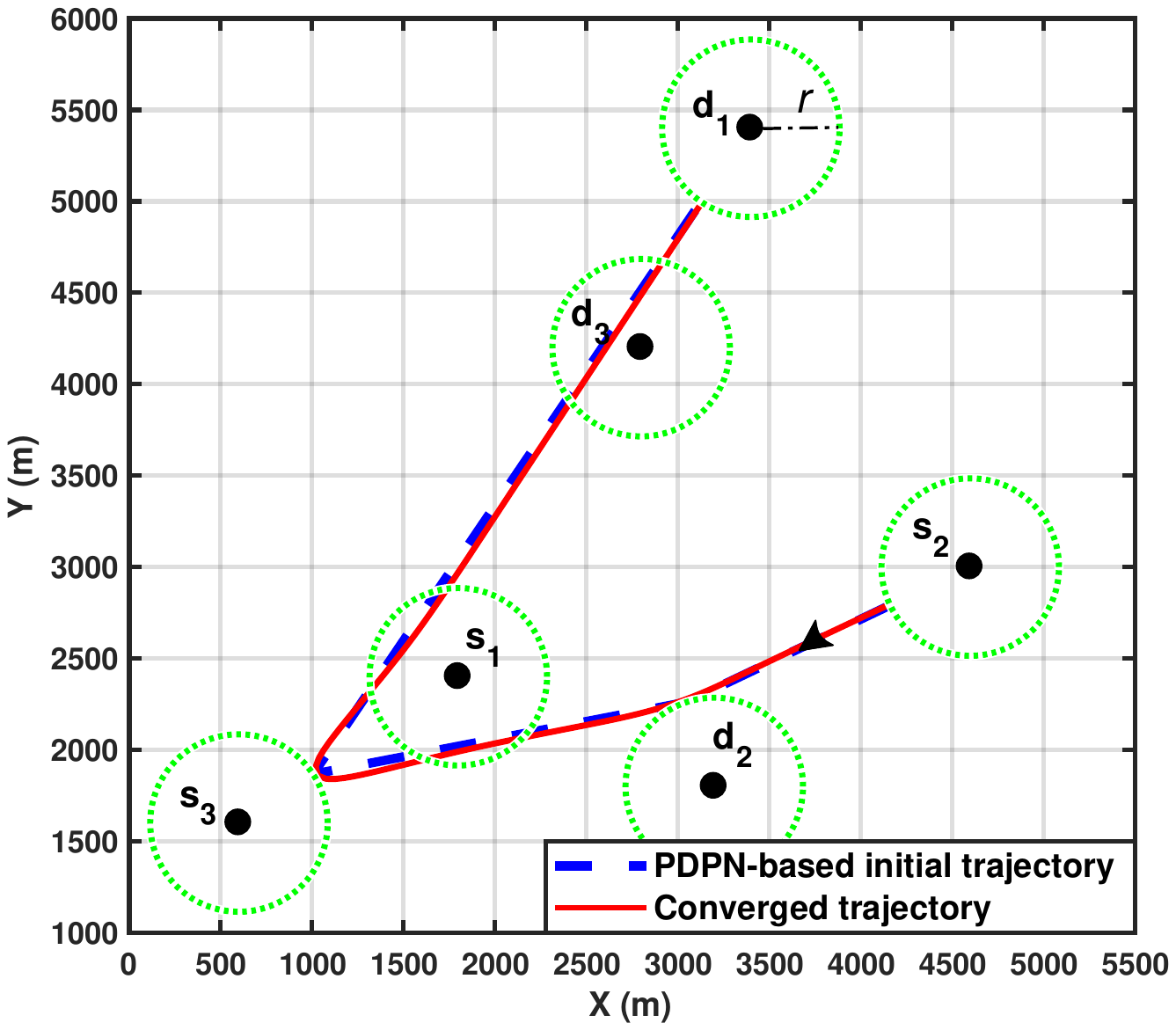}
\vspace{-4.7cm}
\subcaption{\label{300p}PDPN-based initialization ($r=485$ m).}
\endminipage\hfill
\vspace{-0.2cm}
\caption{\label{300one}UAV trajectories with different initializations with throughput requirement $C=300$ Mbits for one-time operation. }
\vspace{-0.5cm}
\end{figure*}
\begin{figure*}[!htb]
\vspace{-2.8cm}
\minipage{0.32\textwidth}
  \hspace*{-1.9cm}\includegraphics[width=1.65\linewidth]{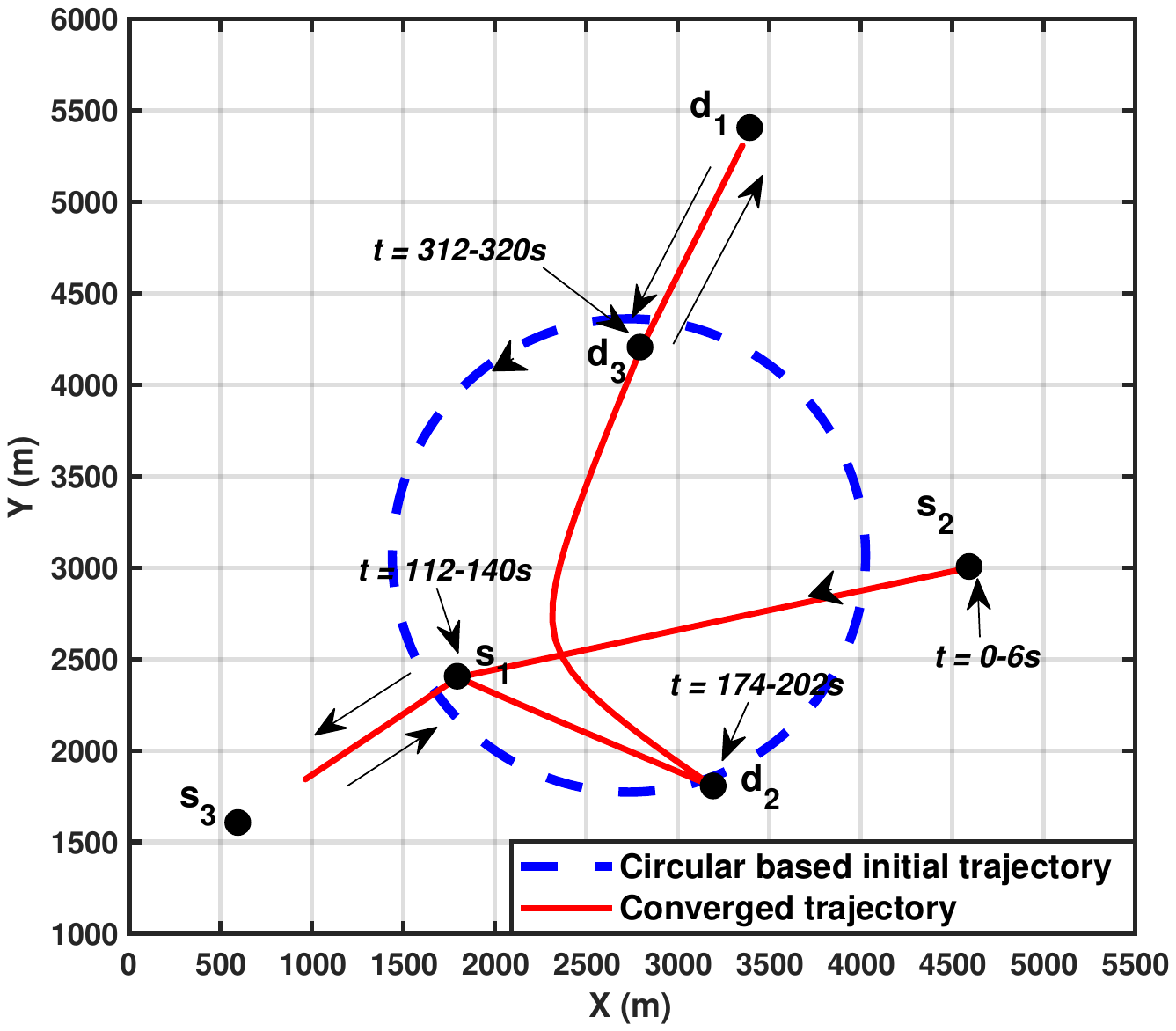}
\vspace{-4.7cm}
  \subcaption{\label{1000cir}Circular based initialization.}
\endminipage\hfill
\minipage{0.32\textwidth}
\hspace*{-1.8cm}\includegraphics[width=1.65\linewidth]{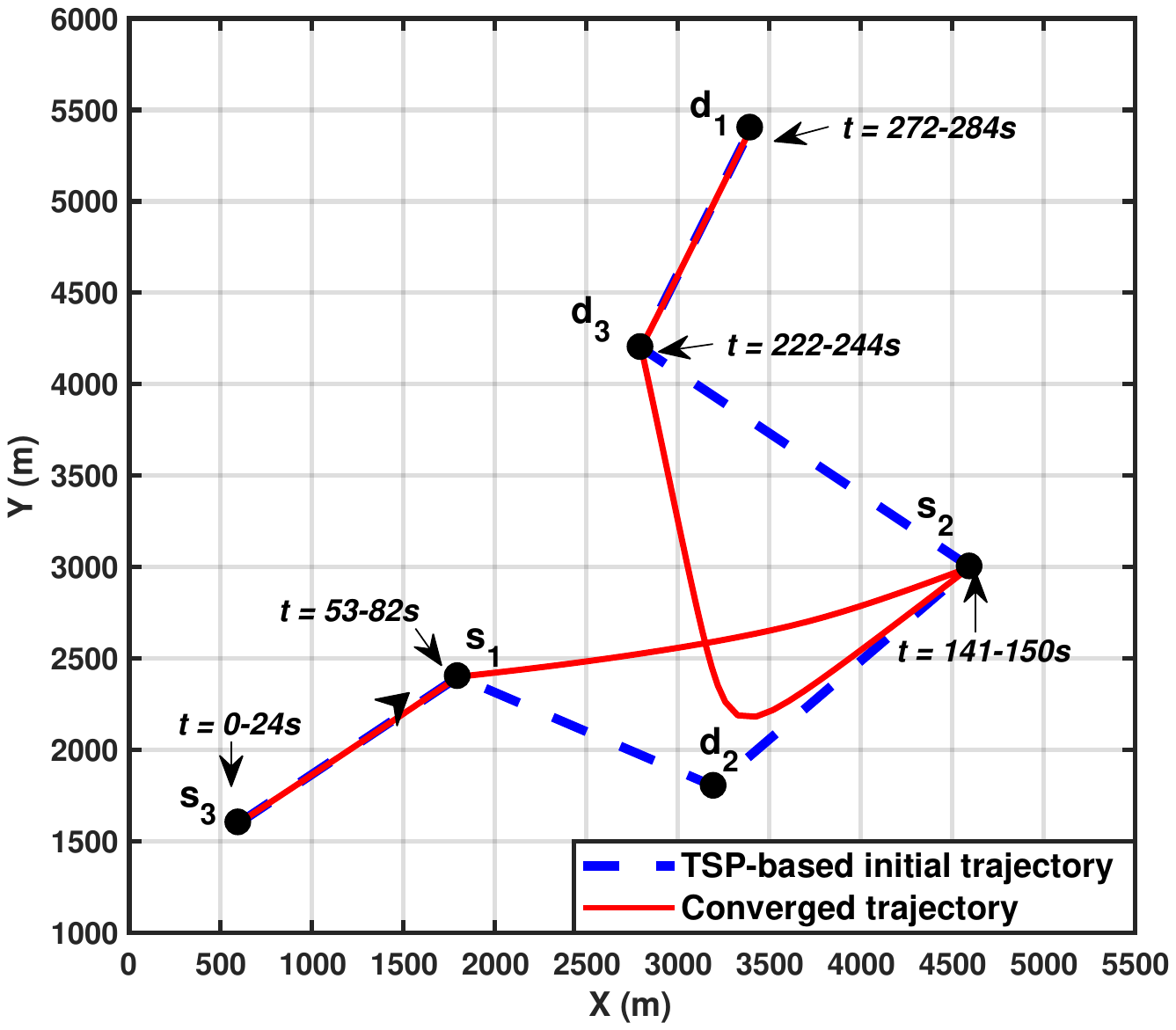}
\vspace{-4.7cm}
  \subcaption{\label{1000t}TSP-based initialization.}
\endminipage\hfill
\minipage{0.32\textwidth}%
 \hspace*{-1.7cm} \includegraphics[width=1.65\linewidth]{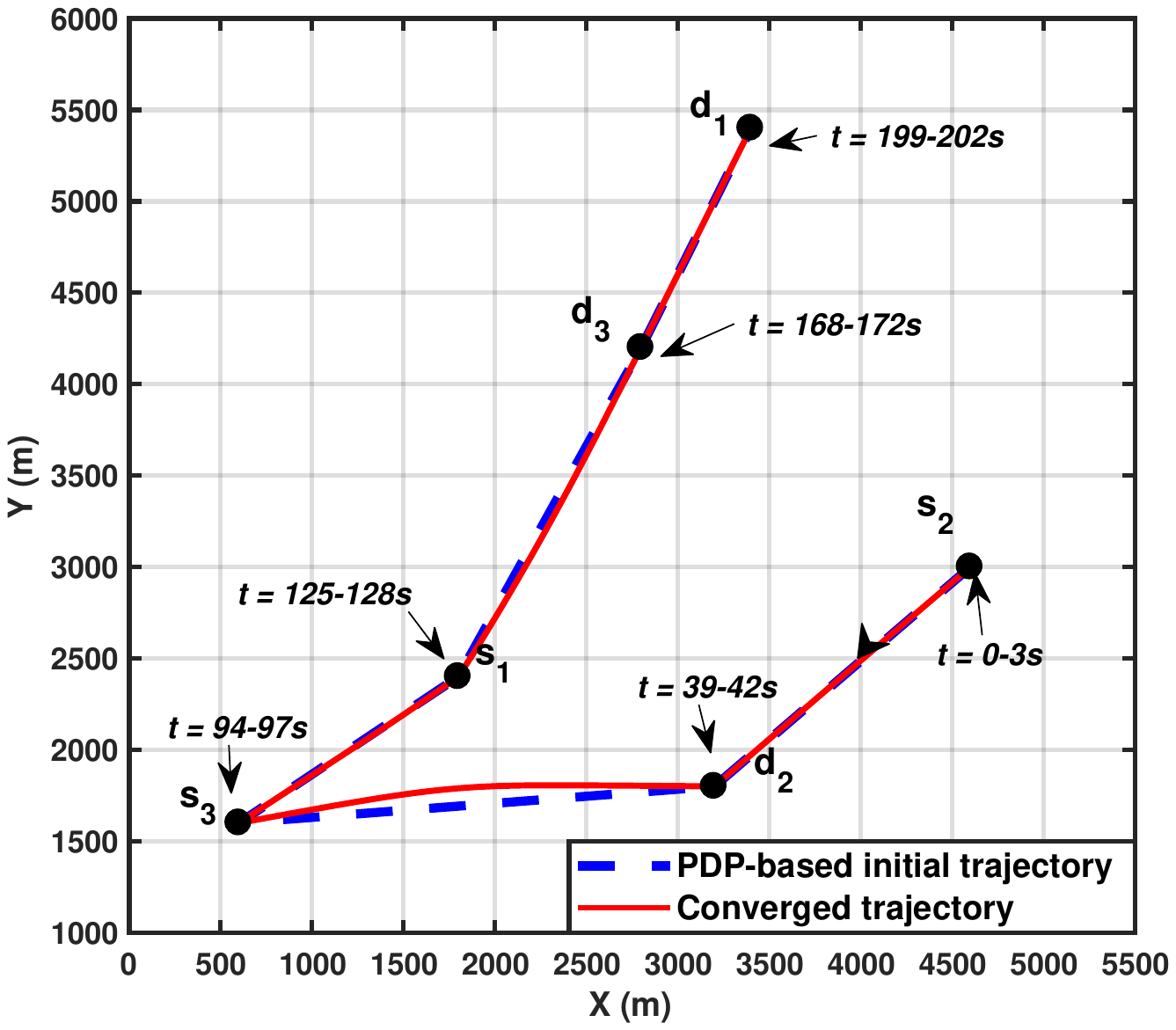}
\vspace{-4.7cm}
  \subcaption{\label{1000p}PDP-based initialization.}
\endminipage\hfill
\vspace{-0.2cm}
\caption{\label{1000one}UAV trajectories with different initializations with throughput requirement $C=1000$~Mbits for one-time operation. }
\vspace{-0.5cm}
\end{figure*}

The minimum flight period for the above two trajectory initialization schemes under different average rate requirements is compared in Fig. \ref{periodcom}. It is observed that at relatively low average rate requirement, the two initialization schemes lead to a comparable performance. This is expected since the UAV is able to finish the mission efficiently even with some moderate link distance from GUs. In contrast, as the average rate requirement increases, the proposed initialization scheme significantly outperforms the circular initialization. This is expected since by explicitly optimizing the visiting order of the GUs, the TSP/TSPN-based initialization ensures that the UAV minimizes its flying time and thus more time can be spent at locations closer to the GUs, which is not attainable by the circular based initialization in general. Furthermore, it is observed that as the flight period $T$ gets sufficiently large, the proposed TSP/TSPN-based initialization approaches the performance upper bound, where each GU communicates with the UAV when the UAV is directly on top of it. The corresponding maximum rate can be calculated as
$\bar{R}^{\mathrm{up}}=B\log_2\left(1+P^v\gamma_0/H^2 \right)/(U+V)\thickapprox 13.86~\mathrm{Mbps}$.

\subsection{One-Time Operation}

In this subsection, we consider the one-time operation scenario as studied in Section \ref{onetime} and Section \ref{inionetime}. For the purpose of exposition, we assume that all source and destination GUs are from Group 3, i.e., $U=V=K_3$. In the following, we compare the required minimum completion time by three different trajectory initializations: 1) circular based trajectory initialization; 2) TSP/TSPN-based trajectory initialization without returning to the initial GU \cite{8255824}; 3) proposed PDP/PDPN-based trajectory initialization in Section~\ref{inionetime}. After solving the corresponding TSP and PDP, the minimum time required to visit all GUs are $T_{\mathrm{tsp}}=166$ s and $T_{\mathrm{pdp}}=186$ s, respectively.

First, with the throughput requirement of $C=300$ Mbits for each source-destination pair, the converged trajectories of the three initializations are plotted in Fig. \ref{300one}. The corresponding minimized completion time is obtained as 150 s, 127 s and 142 s for the circular based, TSPN-based and PDPN-based trajectory initializations, respectively. It is found that the three trajectory initialization schemes lead to different converged trajectories, and the TSPN-based initialization gives the best performance in terms of minimum completion time. This is because with low rate requirement, the UAV is able to finish the mission without having to reach each GU. Therefore, the benefit of the PDPN-based initialization that guarantees approaching the source GUs before destination GUs cannot compensate the longer traveling distance as compared to the TSPN-based initialization.

However, as the throughput requirement increases to $C=1000$ Mbits, the precedence constraints that are considered by the proposed PDP/PDPN-based initialization are expected to make a significant impact, as verified by Fig. \ref{1000one}. The corresponding completion time for the circular based, TSP-based and PDP-based trajectory initializations are respectively $320$ s, $284$ s and $202$ s. It is observed that the three initialization schemes lead to different visiting orders for the GUs. Furthermore, for the circular based initialization in Fig. \ref{1000cir}, it is noted that the UAV fails to reach $\textbf{s}_3$ and $\textbf{d}_1$, and the GUs $\textbf{s}_1$ and $\textbf{d}_3$ are visited twice, which cause unnecessarily longer traveling distance. On the other hand, for the TSP-based initialization shown in Fig. \ref{1000t}, the UAV detours its path towards $\textbf{s}_2$ to collect information before approaching  $\textbf{d}_2$ in the converged trajectory, and thus more time is needed. Besides, since the UAV may reach destination GUs before source GUs in circular based and TSP-based initializations, more power and bandwidth need to be allocated to GUs far from the UAV to satisfy the {\it information-causality constraints}, which results in low spectral efficiency in general. In contrast, for the PDP-based initial trajectory with the precedence constraints considered, even though the initial trajectory has a longer traveling distance compared to that in the TSP-based initialization, it allows the UAV to visit all source GUs before destination GUs such that the power and bandwidth can be more efficiently allocated to the GU that the UAV is hovering above, and hence results in higher spectral efficiency and smaller completion time.

\begin{figure}
\vspace{-3.8cm}
\hspace{-1cm}\includegraphics[width=4.5in]{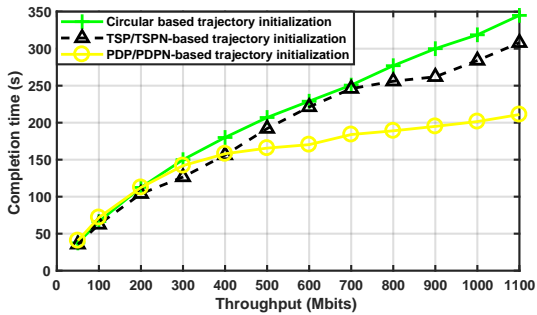}
\vspace{-7.2cm}
\caption{\label{onetim_com}Completion time versus throughput requirement for one-time operation. }
\vspace{-0.5cm}
\end{figure}

For the various initialization schemes, Fig. \ref{onetim_com} shows the required mission completion time versus the throughput requirement. It is observed that for relatively low throughout requirement, all three initializations have a comparable performance, with TSP/TSPN-based trajectory initialization slightly outperforming the PDP/PDPN-based initialization. This is expected, since with a low throughput requirement, the UAV is able to complete the mission without the need to reach each GU so that the precedence constraints are not important. In this case, the benefit of the shortest traveling distance resulted by the TSP/TSPN-based initialization dominates the precedence constraints.

With high throughput requirement, the UAV needs~to reach each GU to enjoy the best communication link. The proposed PDP/PDPN-based trajectory initialization significantly outperforms the other two. This is expected since the power and bandwidth with the PDP/PDPN-based trajectory can be more efficiently utilized to satisfy the {\it information-causality constraints}. Moreover, in this scenario, the proposed algorithm also approaches the performance upper bound, with the required mission completion time increasing linearly with the throughput requirement at sufficiently large $T$. Specifically, when the throughput increases by 100 Mbits, the increase of the completion time is approximately equal to $\Delta T=(U+V)\times 100~\mathrm{Mbits}/\left(B \log_2(1+P^v\gamma_0/H^2 ) \right)\approx7.2$ s.

\section{Conclusion}
\label{conclusion}

This paper studies a general UAV-enabled RAN with multi-mode communications. We consider two UAV operation scenarios of practical interest, namely periodic operation versus one-time operation, for which we formulate and solve the optimization problems to jointly design the UAV trajectory and communication resource allocation to minimize the UAV flight time. We propose iterative algorithms by employing successive convex optimization and block coordinate descent techniques to find efficient locally optimal solutions. Furthermore, we design a TSP/TSPN-based initial trajectory and a PDP/PDPN-based initial trajectory for the UAV in the two operation scenarios, respectively. Numerical results show that significant UAV flight time saving and user throughput improvement are achieved by the proposed trajectory designs compared to that with the existing circular trajectory for initialization. The results of this work can be extended to other practical cases such as that with multiple UAVs \cite{8247211}, moving GUs \cite{liu2018comp}, multiple antennas \cite{6648625}, and/or existing ground BSs \cite{8329013}, as well as that by taking into account the UAV energy consumption in the trajectory design \cite{zeng2017energy}, \cite{zeng2018energy}, which will be left for future work.

\appendices
\section{Proof of Lemma \ref{lemma001}}
\label{proflema01}

For any given period $T$ and any infinitesimal positive quantity $\epsilon$, the corresponding optimal values obtained in (P1.1) are denoted as $\eta^*(T)$ and $\eta^*(T+\epsilon)$, respectively. To prove Lemma \ref{lemma001}, we only need to show that $\eta^*(T)\leq \eta^*(T+\epsilon)$. Note that since the flight period $T$ appears in both the denominator and the integration upper limit on the LHS of \eqref{p11001} and \eqref{p11002}, the proof of such inequality is not obvious. A constructive proof is given below.

Specifically, for the given period $T$, denote the optimal solution to (P1.1) as $\textbf{q}^*(t)$, $p_j^*(t)$, $\alpha_i^*(t)$ and $\beta_j^*(t)$, $t \in [0,T]$. As the period increases to $T+\epsilon$ so that the time interval becomes $t' \in [0, T+\epsilon]$, a one-to-one mapping between $t$ and $t'$ can be obtained by the linear scaling $t'=t(T+\epsilon)/T$, $0\leq t'\leq T+\epsilon$. In this case, a feasible solution to (P1.1) with period $T+\epsilon$ can be constructed by letting  $\tilde{\textbf{q}}(t')=\textbf{q}^*(t)$, $\tilde{\alpha}_i(t')=\alpha^*_i(t)$, $\tilde{\beta}_j(t')=\beta^*_j(t)$, and $\tilde{p}_j(t')=p^*_j(t)$, with $t=t'T/(T+\epsilon)$. It is not difficult to see that with such a construction, all constraints in \eqref{p1003}-\eqref{p1009} are satisfied. Furthermore, the LHS of \eqref{p11001} and \eqref{p11002} satisfy
\begin{align}
\frac{B}{(T+\epsilon)\bar{R}_i^u}\int_0^{T+\epsilon}\tilde{\alpha}_i(t') \log_2 \left(1+\frac{P_i^u \gamma_0}{\tilde{\alpha}_i(t')(H^2+||\tilde{\textbf{q}}\textbf{}(t')-\textbf{s}_i||^2)  } \right) \mathrm{d} t' \nonumber\\
\stackrel{(e)}{=}\frac{B}{T\bar{R}_i^u} \int_0^T\alpha^*_i(t)\log_2 \left(1+\frac{P_i^u\gamma_0}{\alpha^*_i(t)(H^2+||\textbf{q}^*(t)-\textbf{s}_i ||^2)}   \right)\mathrm{d}t \nonumber~~~~~~~ \nonumber
\end{align}
\vspace{-0.6cm}
\begin{align}
\!\!\!\!\geq  \eta^*(T), \ \ \forall i,~~~~~~~~~~~~~~~~~~~~~~~~~~~~~~~~~~~~~~~~~~~~~~~
\end{align}
\vspace{-0.7cm}
\begin{align}
\frac{B}{(T+\epsilon)\bar{R}_j^v} \int_0^{T+\epsilon}  \tilde{\beta}_j(t') \log_2 \left(1+\frac{\tilde{p}_j(t')\gamma_0}{\tilde{\beta}_j(t')(H^2+||\textbf{d}_j-\tilde{\textbf{q}}(t')||^2)}  \right)\mathrm{d}t'    \nonumber\\
\stackrel{(f)}{=}\frac{B}{T\bar{R}_j^v}\int_0^T \beta^*_j(t) \log_2 \left(1+\frac{p^*_j(t)\gamma_0}{\beta^*_j(t) (H^2+||\textbf{d}_j-\textbf{q}^*(t)||^2)} \right)\mathrm{d}t \!\!\! \nonumber ~~~~~~~~
\end{align}
\vspace{-0.5cm}
\begin{align}
\!\!\!\geq  \eta^*(T), \ \ \forall j,~~~~~~~~~~~~~~~~~~~~~~~~~~~~~~~~~~~~~~~~~~~~~~
\end{align}
where both $(e)$ and $(f)$ hold due to the linear transformation of $t'=t(T+\epsilon)/T$.

As a result, based on the optimal solution to (P1.1) with period $T$, we have constructed a feasible solution to (P1.1) with period $T+\epsilon$ that achieves an objective value no smaller than $\eta^*(T)$, which serves as a lower bound for the optimal value $\eta^*(T+\epsilon)$. Therefore, we have $\eta^*(T)\leq \eta^{*}(T+\epsilon)$. This thus completes the proof.

\section{Proof of Proposition \ref{theorem01}}
\label{proftheorem01}

The proof of Proposition \ref{theorem01} is similar to that of Lemma 2 in \cite{7572068} and Theorem 2 in \cite{zeng2017energy}. We first introduce the following function $f(z)\triangleq \log_2\left(1+\frac{\gamma}{\tau+z} \right)$ for some constant $\gamma\geq 0$ and $\tau$, which can be shown to be convex with respect to $z\geq -\tau$. Using the property that the first-order Taylor approximation of a convex function is a global under-estimator \cite{boyd2004convex}, for any given $z_0$, we have $f(z)\geq f(z_0)+f'(z_0)\left(z-z_0 \right)$, $\forall z$, where $f'(z_0)=\frac{-(\log_2e)\gamma}{\left(\tau+z_0\right)\left(\tau+\gamma+z_0\right)}$ is the derivative of $f(z)$ at point $z_0$. By letting $z_0=0$, we have the following inequality
\begin{align}
\!\!\!\!\log_2\left(1+\frac{\gamma}{\tau+z} \right) \geq \log_2\left(1+\frac{\gamma}{\tau}\right)-\frac{(\log_2e)\gamma z}{\tau(\tau+\gamma)}, \ \, \forall z.
\end{align}
Then by letting $\gamma=\varepsilon_{i}[n]$, $\tau=H^2+||\textbf{q}^l[n]- \textbf{s}_i||^2$, and $z=||\textbf{q}[n]-\textbf{s}_i||^2-||\textbf{q}^l[n]- \textbf{s}_i||^2$, we have
 \begin{align}
\phi_{i}^{l}[n] \triangleq\frac{  \alpha_{i}[n] \left( \log_2e\right)\varepsilon_{i}[n]}{\left(H^2+||\textbf{{q}}^l[n]-\textbf{{s}}_i ||^2   \right)\left(H^2+||\textbf{{q}}^l[n]-\textbf{{s}}_i||^2+  \varepsilon_{i}[n] \right)}.~\!
\end{align}
And the inequality \eqref{ruk} can be obtained accordingly.

Similar results can also be obtained for $R^v_{j}[n]$ and $\hat{R}_{j}^{v}[n]$ in \eqref{rdk}, and $\varphi_{j}^l[n]$ can be defined as \begin{align}
\varphi_{j}^l[n]\triangleq\frac{  \beta_{j}[n]\left(\log_2e\right)\zeta_{j}[n]}{(H^2+||{\textbf{d}}_j-\textbf{{q}}^l[n] ||^2)(H^2+||{\textbf{d}}_j-\textbf{{q}}^l[n] ||^2+\zeta_{j}[n] ) }.  \!
\end{align}
The details are omitted for brevity.

\section{Overview of Pickup-and-Delivery Problem}
\label{over_pdp}
In this section, we give a brief overview of the classic PDP, which is also known as DARP \cite{parragh2008survey}, \cite{cordeau2003dial}. In the most basic form of PDP and DARP, a capacitated vehicle must satisfy a set of transportation requests, where each request specifies an origin (pickup point) and a destination (drop-off point). The objective is to design a minimum-cost vehicle route accommodating pairing and the following precedence constraints: for each request, the origin must be visited before the destination. The difference between PDP and DARP is that PDP usually deals with problems like goods transportation, while DARP refers to passenger delivery with additional constraints involved. Both PDP and DARP are generalizations of the classic TSP and thus NP-hard. Various heuristic and approximation algorithms have been proposed to yield good results within a reasonable time complexity. By employing the CPLEX CP Optimizer \cite{WinNT}, an optimal result can be obtained for small-size problems efficiently.

\ifCLASSOPTIONcaptionsoff
  \newpage
\fi

\bibliographystyle{IEEEtran}
\bibliography{reference/ref}

\end{document}